\documentclass[10pt]{emulateapj}
\usepackage{apjfonts}

\usepackage{natbib}
\usepackage[usenames]{color}
\newcommand{\beq}{\begin{equation}} 
\newcommand{\eeq}{\end{equation}} 
\newcommand{\beqa}{\begin{eqnarray}} 
\newcommand{\eeqa}{\end{eqnarray}}

\newcommand{\pav}{$\bar{\mathrm{P}}_{\rho\mathrm{12}}$\ }
\newcommand{\pavm}{$\bar{\mathrm{P}}_{\mathrm{M}}$\ }
\newcommand{\rtwelve}{10$^{12}$ g cm$^{-3}$}

\def\mo{M$_\odot$}

\def\Dwa{$\,$\uppercase\expandafter{\romannumeral5}$\,$}

\def\sless{\lower2pt\hbox{$\buildrel {\scriptstyle <}
   \over {\scriptstyle\sim}$}}

\def\sgreat{\lower2pt\hbox{$\buildrel {\scriptstyle >}
   \over {\scriptstyle\sim}$}}
\def\sharpnull#1{}

\begin{document}

\slugcomment{Submitted to ApJ. August 22, 2005. Accepted December 22, 2005}

\title{The Spin Periods and Rotational Profiles of Neutron Stars
at Birth\vspace*{.5cm}}

\author{Christian D. Ott}
\affil{Max-Planck-Institut f\"{u}r Gravitationsphysik,
Albert-Einstein-Institut, Golm/Potsdam, Germany; cott@aei.mpg.de}
\author{Adam Burrows}
\affil{Department of Astronomy and Steward Observatory, 
The University of Arizona, Tucson, AZ 85721;
burrows@zenith.as.arizona.edu}
\author{Todd A. Thompson\footnote{Hubble Fellow}}
\affil{Current Addres: Department of Astrophysical Sciences, Princeton
   University, Peyton Hall - Ivy Lane Princeton,
   NJ 08544; thomp@astro.princeton.edu}
\author{Eli Livne}
\affil{Racah Institute of Physics, The Hebrew University,
Jerusalem, Israel; eli@frodo.fiz.huji.ac.il}
\author{Rolf Walder}
\affil{Department of Astronomy and Steward Observatory, 
The University of Arizona, Tucson, AZ 85721;
rwalder@as.arizona.edu}

\begin{abstract}
We present results from an extensive set 
of one- and two-dimensional radiation-hydrodynamic simulations of 
the supernova core collapse, bounce, and postbounce phases, and 
focus on the protoneutron star (PNS) spin periods and rotational profiles 
as a function of initial iron core angular velocity, 
degree of differential rotation, and progenitor mass. 
For the models considered, we find a roughly linear mapping between initial
iron core rotation rate and PNS spin. The results
indicate that the magnitude of the precollapse iron core angular
velocities is the single most important factor in determining the
PNS spin. Differences in progenitor mass and 
degree of differential rotation lead only to small variations in the
PNS rotational period and profile.
Based on our calculated PNS spins, 
at $\sim$200-300 milliseconds after bounce, and assuming 
angular momentum conservation, we estimate final
neutron star rotation periods. We find periods of one millisecond
and shorter for initial central iron core periods of below $\sim$10~s.
This is appreciably shorter than what previous studies have predicted
and is in disagreement with current observational data from 
pulsar astronomy. After considering possible spindown mechanisms 
that could lead to longer periods we conclude that there is 
no mechanism that can robustly spin down a neutron star from $\sim$1~ms
periods to the ``injection" periods of tens to hundreds of 
milliseconds observed for young pulsars. 
Our results indicate that, 
given current knowledge of the limitations of neutron star spindown 
mechanisms, precollapse iron cores
must rotate with periods around 50-100~seconds to form neutron stars 
with periods generically near those inferred for 
the radio pulsar population.
\end{abstract}

\keywords{Stars: Rotation, Stars: Evolution, Stars: Neutron, Pulsars:
 General, Stars: Supernovae: General, 
 Hydrodynamics, Neutrinos, Radiative Transfer}

\section{Introduction}
\label{section:intro}
Neutron stars are formed in the core 
collapse of dying massive stars or (rarely) 
in the accretion-induced collapse (AIC) of O-Ne-Mg white 
dwarfs in binary systems. 
In both types of collapse events an electron-degenerate core
is pushed over its effective Chandrasekhar mass.
The collapse is halted only when 
the core density reaches $\sim$2 $\times$ 10$^{14}$ g cm$^{-3}$ 
and nuclei undergo a phase transition to free nucleons. The
equation of state (EOS) abruptly stiffens at that point and
the core bounces, launching a hydrodynamic shock wave into the
supersonically infalling outer core and leaving behind a
hot and extended protoneutron star (PNS). The
bounce shock stalls almost immediately and turns into an
accretion shock due to a combination
of neutrino losses, the ram pressure of the infalling mantle,
and the dissociation of heavy nuclei into neutrons, protons and
$\alpha$ particles. 
The subsequent evolution of the system
to a potentially successful supernova explosion has been under 
detailed investigation for many decades now. The favored
``delayed neutrino mechanism'' (\citealt{bethewilson:85}) relies
on shock revival due to sufficient energy deposition
by neutrinos in the so-called ``gain region'' behind the stalled
shock. However, using numerical simulations that 
incorporate Boltzmann neutrino transfer and the best 
available physics, a number of groups have shown that this mechanism
does not succeed in spherical symmetry
(\citealt{tonyetal1:01,liebendoerfer:01,rampp_janka,thompetal:03}). In spherical
symmetry, the bounce shock is not revived, though an increase of 
only~$\sim$25\% in the net energy deposition 
(all else being equal) would be sufficient 
to drive a successful explosion
(\citealt{thompvisc:05}).  However, recent two-dimensional 
simulations (\citealt{herant:94};
\citealt{bhf:95}; \citealt{fryerheger:00}; \citealt{jankaetal:03}; 
\citealt{burasetal:03};
\citealt{walder:05}), and the first three-dimensional  
simulations (\citealt{fryerwarren:04}; \citealt{jankaetal:05}) indicate
that indeed multi-dimensional phenomena such as postbounce 
convection and/or rotation might lead to the 
conditions needed for a successful explosion. 
In the same context,
Thompson et al.~(2005) performed simulations in spherical symmetry,
including a 1D prescription for rotation, angular momentum transport,
and viscous heating.  A number of their more rapidly rotating models 
exploded due to the additional energy deposited behind the stalled 
shock by viscous dissipation.

It is, however, possible that only stars of
masses less than $\sim$25-30~\mo\ explode as supernovae, 
leaving neutron stars (\citealt{fryer:99,hegeretal:03}).
The fate of more massive progenitor stars is more uncertain and
may be associated with under-energetic supernovae 
(\citealt{nomoto2005}), hypernovae, gamma-ray bursts (GRBs), and/or 
black hole formation (\citealt{fryer:99,fryerkalogera:01,
hegeretal:03}).   
If shock revival and a strong supernova explosion 
result, no critical amount of mass is accreted onto the PNS. 
The PNS cools and contracts quasi-hydrostatically  
to its final radius of $\sim$10~-~15~km on a 
timescale of {\it tens\rm} of seconds, radiating away 
$\sim3\times 10^{53}$~ergs in neutrinos (\citealt{bl:86}; 
\citealt{keiljanka:95}; \citealt{ponsetal:99}). 

Long after the supernova explosion,  
the still young neutron star becomes visible to observers, possibly
as a canonical rotation-powered radio pulsar
 (\citealt{kaspihelfand:02}). Observationally, 
it is often difficult to directly associate pulsars with 
supernova remnants (SNRs).  This is partly due to the
difference between the lifetimes of a typical pulsar
(10$^{6-7}$ years) and a SNR (10$^{4-5}$ years) (\citealt{lyne:98}).
In addition, pulsars have observed average space velocities
of 200 - 500 km s$^{-1}$ (\citealt{lynelorimer:94}), most probably 
imparted to them during the supernova. These ``pulsar kicks''
may be a strong indication of the multi-dimensional nature of the 
supernova mechanism (\citealt{bh:96}; \citealt{laikick:01};
\citealt{fryer:04}; \citealt{scheck:04}).

Observations of radio pulsars imply that the initial rotation
rate of neutron stars is in the vicinity of tens to
hundreds of milliseconds (\citealt{kaspihelfand:02}). The fastest
known young pulsar is PSR~J0537-6910 which is associated
with SNR~N157B and has a spin period of 16 ms and an 
age of $\sim$5~kyr. If one assumes a pulsar breaking index, n, less than
or equal to 2
\footnote{A breaking index of 3 holds for pure magnetic dipole 
radiation and n~<~2-3 is common
for young pulsars (\citealt{lyne:98}).}, an initial pulsar
period  \sless\ 10 ms is obtained 
(\citealt{marshall:98}). For the Crab pulsar (PSR B0531+21, P = 33 ms),
n has been measured and because the year of the Crab's birth is
known independently (1054 A.D.), a solid initial spin determination of 
P$_{\mathrm{initial}}$~=~19~ms has been made 
(\citealt{kaspihelfand:02}).

The empirically determined initial spin periods of pulsars have to be
matched to the theoretical models of stellar evolution, 
core collapse, supernova explosion,
and neutron star cooling. Massive stars are suspected to rotate
rigidly and rapidly during hydrogen burning on the main sequence. 
Their equatorial velocities are typically $\sim$200~km~s$^{-1}$, which 
corresponds to $\sim$10\% of their Keplerian break up 
speed (\citealt{fukuda:82}). Since there is no detailed empirical
knowledge of the evolution of the angular momentum distribution
inside an evolving star and because consistent 
multi-dimensional stellar evolutionary calculations have yet 
to be performed, stellar evolution theorists
must rely on parameter-dependent, semi-phenomenological prescriptions
to follow the angular momentum evolution of massive stars in their 
spherically symmetric models. \cite{heger:00} have presented 
 the first rotating presupernova stellar models that included 
a 1D prescription for angular momentum transport and centrifugal effects
(the latter only up to the end of core carbon burning). These authors 
estimated the neutron star spin rate by assuming that the
total angular momentum contained in the precollapse iron core 
is conserved during collapse and supernova phases  
and is deposited completely in a rigidly rotating neutron star of 
radius and moment of inertia (I) of
$\sim$12~km and ~0.35~M$_{\mathrm{NS}}$~R$_\mathrm{NS}^2$, 
respectively
(\citealt{lattimerprakash:01}). 
In this simple way, they estimated
an initial neutron star period of $\sim$1~ms for their 
fiducial 20 \mo\ (ZAMS) stellar model. Given the observational
data, this initial period is at least a factor 
of ten too short for the generic pulsar. 
Processes that could lead to an early spindown of a young 
(proto)neutron star include:
angular momentum redistribution by
global hydrodynamic rotational instabilities 
(\citealt{cent:01}; \citealt{rotinst:05}), r-modes and 
gravitational radiation back-reaction 
(\citealt{holai:00}; \citealt{lindblometal:01}; \citealt{watts:02}; 
\citealt{arras:03}), 
rotation-powered explosions (\citealt{akiyama:03}; 
\citealt{akiyama:05}),
viscous angular momentum transport due to
convection, neutrino viscosity or dissipation of shear
energy stored in differential rotation (\citealt{thompvisc:05}),
magneto-centrifugal winds (\citealt{thomp:03}; 
\citealt{thompmagnetar:04}), early magnetic-dipole radiation 
(\citealt{ostrikergunn:69}), late-time fall-back 
(\citealt{heger:04}), and anisotropic 
neutrino emission (\citealt{jankaproc:04}).

It is also possible that slowly spinning cores are the natural
 end-product of stellar evolution.
\cite{spruitphinney:98} included angular momentum 
transport via magnetic processes that yielded such large neutron star 
periods (P~$\sim$100 s) that the authors had to rely on 
subsequent spin-up 
by off-center birth kicks to obtain spin periods matching the 
observations. With a similar approach for magnetic angular momentum 
transport during stellar evolution and the simple neutron star spin 
estimate described above, 
\cite{hegerwoosleyspruitlanger:03} and \cite{heger:04} obtained 
neutron star periods of $\sim$10~ms. However, given the differences in 
estimated iron core spin rates and angular momentum profiles 
in the literature and keeping in mind that the study of stellar 
evolution with rotation is
still in its infancy, one should be cautious in accepting any of the 
recent estimates as a final answer. Too many open issues remain and 
can only be resolved by truly multi-dimensional stellar evolution 
studies in the future. Until then, rotating 1D evolutionary models 
are the only means we have of estimating the angular momentum 
distribution and spin rate 
of iron cores at the onset of core collapse. 
During the subsequent collapse 
evolution, the core spins up due to angular momentum conservation and 
--- depending upon its initial distribution and amount of 
angular momentum ---  becomes oblate due to rotational flattening. 
If the rotating core deviates significantly from spherical symmetry, 
a multi-dimensional treatment becomes indispensable to obtain reliable 
estimates of neutron star birth periods.

To date, only a small number of multi-dimensional supernova studies 
have addressed the neutron star spin question. 
\cite{fryerheger:00} performed 
smooth-particle hydrodynamics simulations with a gray, 
flux-limited diffusion scheme for radiative transfer of 
the neutrinos. They used the rotating progenitor 
model E15B of \cite{heger:00} and estimated an initial spin period of
 the PNS  (at the end of their simulation and assuming 
rigid rotation) on the order of 100~ms. It is, however, not clear how 
they defined the extent of the PNS. Taking the angular 
momentum contained in their PNS 
and following \cite{heger:00} they obtained a neutron star 
spin period of $\sim$2~ms, even for a progenitor model that included 
an $\alpha$-disk-like prescription for turbulent-viscous angular 
momentum redistribution. In their recent gray 3D smooth-particle 
radiation-hydrodynamics calculation,
 \cite{fryerwarren:04} estimated neutron star spin periods by assuming 
that the angular momentum of the inner 1~\mo\ is conserved as the 
PNS cools and contracts to a neutron star. For model E15A 
(\citealt{heger:00}) they estimated in that way a neutron star 
period of 0.91~ms and for the slow ``magnetic'' 15~\mo\ model of 
\cite{hegerwoosleyspruitlanger:03} a period of 17~ms. 
\cite{ott:04} have performed purely hydrodynamic simulations 
with a detailed nuclear equation of state and realistic progenitor 
models and focused on the impact of rotation on the gravitational wave 
signatures of stellar core collapse. 
They argue that if gravitational waves of a core-collapse event were 
observed with sufficient precision it would be possible to infer 
details about the rotational configuration from the observations. 

Using the code VULCAN/2D, \cite{walder:05} performed 
a series of 2D multi-group, multi-species 
flux-limited diffusion radiation-hydrodynamics supernova 
simulations with the 11~\mo\ (ZAMS) progenitor
model of \cite{ww:95}, which they artificially forced to rotate 
according to a rotation law that enforces constant angular velocity on 
cylindrical shells. Their focus was on the anisotropies in the 
neutrino-radiation field and in neutrino heating due to the rotational 
flattening of the postbounce core and they did not systematically 
analyze the final rotational configuration of their PNSs.
In their 1D supernova simulations, \cite{thompvisc:05} included
the effects of rotation approximately and they studied the action
of viscous processes in dissipating the strong rotational 
shear profile produced
by core collapse in a range of progenitors and for different initial
iron core periods.  They showed that for rapidly rotating cores
with postbounce periods of~\sless~4~ms,
viscosity (presumably by magnetic torques, e.g. the magneto-rotational
instability [MRI] or magneto-convection) can spin down the rapidly rotating 
PNSs by a factor of $\sim2-3$ in the early 
post-bounce epoch.

This paper is the first in a series of papers in which we plan to make  
progressively better estimates of neutron star birth periods 
and rotational configurations, connecting supernova theory 
with pulsar science.
Here, we present the first systematic investigation
of the mapping between initial iron core spin and the rotational
configuration of neutron stars at birth carried out with 
state-of-the-art one- and two-dimensional radiation-hydrodynamics
codes. We include in our model suite the nonrotating 11\mo,
15\mo\ and 20\mo\ progenitor models of \cite{ww:95}, the 
rotating 15\mo\ and 20\mo\ progenitors of \cite{heger:00} and the recent
15\mo\ rotating progenitor of \cite{heger:04} that has been evolved
with angular momentum redistribution by magnetic fields. We
systematically investigate the dependence of the PNS
spin on progenitor mass and structure and compare the PNS rotational
configurations resulting from the nonrotating progenitors which 
we force to rotate according to a prescribed rotation law to those 
resulting from models that have been evolved with 
a one-dimensional treatment of rotation.

In \S\ref{section:methods}, we describe the supernova codes
SESAME and VULCAN/2D which we employ in this study. 
Section~\ref{section:initialmodels} reviews the progenitor model
suite on which we are relying and discusses the 
rotation laws and mappings used. Section \ref{section:results} 
describes our calculations in one and two spatial dimensions. 
In \S\ref{section:comp1D2D}, we compare one-dimensional
and two-dimensional simulations and assess the quality of the
rotating one-dimensional models. In \S\ref{section:simpleP}, 
based on the PNS structures that we find in our 
simulations, we estimate the final cold neutron star spin
and discuss spindown mechanisms. Finally, in \S\ref{section:sumdisc}
we summarize our results and discuss them critically.

\section{Computational Methods}
\label{section:methods}

\subsection{1-D, Spherically Symmetric Simulations}
\label{section:1Dmethods}

Our spherically symmetric models are computed with the 
radiation-hydrodynamics algorithm 
SESAME\footnote{{\bf S}pherical {\bf E}xplicit/Implicit 
{\bf S}upernova {\bf A}lgorithm for {\bf M}ulti-Group/Multi-Angle 
{\bf E}xplosion Simulations.}, described in 
detail in \cite{burrowsetal:00} and \cite{thompetal:03}. 
The hydrodynamics scheme is Lagrangean, Newtonian, and
explicit. It is coupled to the EOS of \cite{lseos:91} (the LSEOS), 
using the implementation described in \cite{thompetal:03}.
The comoving Boltzmann equation for neutrinos is solved implicitly
and to order $v/c$ (where $v$ is the matter velocity).

\cite{thompvisc:05} have extended SESAME to approximately include the 
effects of rotation, as well as angular momentum transport
by viscosity and local viscous heating.  The strong negative shear profile
($d\Omega/d\ln r$) 
generated by core collapse implies that much of the region interior
to the supernova shock is unstable by the generalized 
Solberg-H\o iland criterion and, thus, to the magneto-rotational instability 
(MRI; Balbus \& Hawley 1991, 1994; see also Akiyama et al.~2003).  
Under the assumption that magnetic torques 
provide the dominant viscosity in the PNS, by equating
the shear and Maxwell stresses one derives that the shear viscosity
should scale as $\xi\sim h^2 \Omega$, where $h$ is the pressure scale 
height.  Following Thompson et al.~(2005), we introduce an 
$\alpha$-parameter (in analogy with studies of accretion) and 
take $\xi=\alpha h^2 \Omega$.
With this form for the viscosity the viscous heating rate 
scales as $\Omega^3$. The viscous timescale is 
$\tau_{\rm visc}\sim(\alpha\Omega)^{-1}(r/h)^2$.  
If the MRI does operate in PNSs at birth, any seed magnetic 
field grows exponentially on a timescale $\sim\Omega^{-1}$.  
Although the saturation magnetic field strength is somewhat uncertain,
to first approximation the toroidal magnetic field grows until
$B_{\rm sat, \,\phi}\sim(4\pi\rho)^{1/2}r\Omega$, 
which is of order $\sim10^{16}$~G
for $\rho=10^{12}$~g~cm$^{-3}$, $r\sim40$~km, and a protoneutron 
star spin period of $\sim10$~ms just after collapse
(Akiyama et al. 2003).
The poloidal magnetic energy density saturates at a value
approximately an order of magnitude smaller than $B_{\rm sat, \,\phi}^2/8\pi$
(e.g., Balbus \& Hawley 1998).

All rotating 1D models (:1D suffix in the model name) 
are evolved with 500 mass zones, encompassing the 
progenitor iron core. Twenty energy groups are employed in
the radiation transport per neutrino species. Models 
that include viscosity have an  additional ``v" added to 
their model name (see Table~\ref{table:initialmodels}).
All viscous models assume $\alpha=0.1$.  Viscosity is 
turned on at bounce and is assumed to act at all radii
interior to the shock radius.

\subsection{2-D, Axisymmetric Simulations}
\label{section:2Dmethods}
Our axisymmetric core-collapse supernova models are evolved using the 2D 
Newtonian,  multi-species, multi-group, flux-limited (MGFLD) variant of 
the multi-species, multi-group, multi-angle, time-dependent 
radiation-hydrodynamics code VULCAN/2D (\citealt{livne:04}). This
variant has  previously  been used and described in the study 
of \cite{walder:05}. Its hydrodynamics module  
(\citealt{livne:93}) uses an Arbitrary Lagrangean-Eulerian scheme with 
remap and a scalar von Neumann-Richtmyer artificial viscosity 
scheme for shock handling. We employ the LSEOS, using the implementation 
described by \cite{thompetal:03}.

Since VULCAN/2D uses an Eulerian grid, the specific angular momentum
is advected with the flow in the same manner as linear momemtum 
components. In so doing, we maintain global angular momentum
conservation by construction.  Note that the axis in cylindrical 
coordinates is a singularity and, as such, is prone to slightly larger 
errors than can be expected elsewhere on the grid. However, the actual
volume of the cells nearest the axis is very small and the
errors do not affect the overall flow. In the past
(\citealt{walder:05}) we have estimated such departures near the singularity 
for rotating models to be no more than $\sim$10\% in any flow variable, 
and to be much smaller elsewhere. 

Using the MGFLD variant of VULCAN/2D allows us to perform an extensive 
study that encompasses the evolution of many rotating models 
to times up to $\sim$300~ms after core bounce. However, one should 
bear in mind that MGFLD is only an approximation to full Boltzmann
transport and differences with the more exact treatment will emerge 
in the neutrino semi-transparent and transparent regimes above the PNS 
surface. Inside the neutrinosphere of the PNS the 
two-dimensional MGFLD approach provides a very reasonable 
representation of the multi-species, multi-group neutrino radiation 
fields. The flux limiter used in this study 
is a vector version of that of \cite{bruenn:85}, 
as discussed in \cite{walder:05}. In the current version of VULCAN/2D
we leave out velocity terms, such as Doppler shifts, but include advection
in the transport sector. Such velocity terms are deemed to be of relative
importance by the Garching group who find neutrino-driven explosions
when turning them off (\citealt{rampp_janka};
\citealt{burasetal:03},2005). 
Note, however, that velocity terms in Eulerian transport
are different from the corresponding terms in the comoving frame and
that statements about their relative importance are very
frame-dependent. We also do not include energy redistribution
by neutrino-electron scattering that may affect the size of the
homologous core at bounce by $\sim$10\%, but is otherwise 
quite subdominant.

A key feature of VULCAN/2D is its ability to deal with 
arbitrarily shaped grids while using cylindrical coordinates 
in axisymmetry. The grid setup we use for this study is similar
to the one depicted in \cite{ott:04}. All our models are evolved on
the full 180~degrees of the symmetry domain. Those models 
from the simulations of \cite{walder:05} are evolved with 80 
evenly spaced angular zones and 130 logarithmically distributed 
radial zones out to 2000~km. All \cite{walder:05} 
models are evolved with 8 neutrino energy groups for each neutrino 
species. Our previously unpublished models are evolved without 
exception with 16 energy groups per neutrino species, 120 angular and 
220 radial zones out to 3000 km, and, for very massive presupernova 
cores (namely models s20, E15A and E20A; see
\S\ref{section:initialmodels}), 
out to 5000 km. We have performed a resolution comparison simulation
with 2/3 the angular and radial zones for a rapidly rotating model
and find that, despite local qualitative and quantitative differences 
in the flow, integral and global observables, such as the 
moment-of-inertia weighted mean period (see \S\ref{section:results}) 
remain practically unchanged, at least at times up to
$\sim$150~-~200~ms after bounce. 
At the resolution we use in this study, VULCAN/2D conserves energy to
better than $\sim$0.7\% near bounce and $\sim$0.4\% on average in
terms of $\Delta$E/E${_\mathrm{grav}}$. At the end of the simulation
$\Delta$E/E${_\mathrm{grav}}$ is $\sim$0.2\%.

\section{Initial Models}
\label{section:initialmodels}
We use two sets of realistic supernova progenitor models from
stellar evolutionary calculations as initial data
for our simulations. The first set encompasses models s11, 
s15, and s20 (ZAMS mass of 11~\mo\ , 15~\mo\ , 20~\mo; 
solar metallicity) of \cite{ww:95}.
These models were evolved in spherical symmetry until the 
onset of iron core collapse. In Fig.~\ref{fig:initialrhomass},
we compare the matter density profiles 
of the presupernova models. Note that
models s11 and s15 have nearly identical structures out to $\sim$1000~km.
The s20 model has a shallower density profile and a
lower central density, but a  more massive iron core~($\sim$1.9\mo, see 
Table~\ref{table:initialmodels}).

On mapping to our simulation grids
we introduce rotation using the rotation law,
\begin{equation}
\label{eq:rotlaw}
\Omega(r) = \Omega_0 \, 
\bigg[ 1 + \bigg(\frac{r}{{\rm \mathrm{A}}}\bigg)^2 \bigg]^{-1}\, ,
\end{equation}
where $\Omega_0 = 2\pi \mathrm{P}^{-1}_0$ is the angular velocity 
corresponding to the initial central rotation period P$_0$.
A is a free parameter that determines the
degree of differential rotation in our models. For $r$~\sless~A,
the initial rotation profile is roughly solid-body.
According to the Poincar\'{e}-Wavre theorem, degenerate rotating
fluid bodies have constant specific angular momentum on 
{\it cylinders \rm} (see, e.g., \citealt{tassoul:78}). 
Hence, we choose $r$ in 
eq.~(\ref{eq:rotlaw}) to be the 
{\it cylindrical radial coordinate \rm} (``cylindrical rotation'')
of the two-dimensional models that we evolve with VULCAN/2D. 
In spherical symmetry, the approximate treatment of rotation goes 
along with the assumption of constant angular momentum on 
{\it spherical \rm} shells (``shellular rotation'') and we interpret 
$r$ as the \it spherical radial coordinate \rm in all models evolved  
in spherical symmetry with SESAME. To compare ``shellular'' and 
``cylindrical'' rotation laws, one of our two-dimensional models
 is set up to rotate initially with constant angular momentum 
on spherical shells. Figure~\ref{fig:initialrot} demonstrates that the 
above rotation law reasonably approximates the rotational profiles of 
the rotating presupernova models of \cite{heger:00} if one chooses 
A$\sim$1000~km. Except for two models in which we investigate the 
effect of A $\sim\infty$, all other models of this set have  
A=1000~km. We name our models according to the following 
convention:  [progenitor name]A[in km]P[P$_0$ in s][:evolution type].
For example, s11A1000P8.00:1Dv is a \cite{ww:95} 11~\mo~model
with A~=~1000~km and P$_0$~=~8.00~s that was evolved in one-dimension 
with viscous dissipation turned on after core bounce. 
Other evolution types are :1D (one-dimensional, no viscous 
dissipation), :2D (two-dimensional, ``cylindrical rotation''), 
2Ds (two-dimensional, ``shellular rotation'') and 2Dh 
(two-dimensional, ``cylindrical rotation,''
purely hydrodynamic, no neutrino transfer). 
Table \ref{table:initialmodels} details our choices of progenitor 
model, value of A, P$_0$, and evolution type for this model set. 
In particular, we choose 
the range of P$_0$ so that Keplerian rotational velocities are 
not attained anywhere during the numerical evolution, that is 
$\Omega(r,t) < \Omega_{\mathrm{Kepler}} = \sqrt{GM/r^{3}}$
for all $r$ and~$t$. In terms of the widely used 
rotation parameter $\beta$, which is defined as the
ratio of rotational kinetic to gravitational potential
energy (T/$|$W$|$), for this series the precollapse values
range from 7$\times$10$^{-5}$\% up to 0.88\% for the fastest model.

The second set of progenitors (``Heger models'' in the following) 
which we consider consists of models E15A, E20A (\citealt{heger:00}) 
and m15b6 (\citealt{heger:04}). These progenitor models were evolved 
to the onset of iron core collapse with the approximate treatment of 
rotation described by Heger et al. (2000), but centrifugal forces were 
included only until the end of core carbon burning. Model E15A and 
model E20A are relatively fast rotators with initial central periods
of $\sim$1.5 and 2 seconds and values 
of $\beta$ = T/$|$W$|$ of $\sim$0.5\% and 0.3\%, 
respectively. In model m15b6, angular momentum
redistribution by magnetic processes was included. This 
led to slow and more rigid rotation of its iron core,
as can be inferred from Fig.~\ref{fig:initialrot}.
In Fig.~\ref{fig:initialrhomass}, we compare the density  
profiles of the ``Heger models'' with those of the nonrotating
s11, s15, and s20 models. The slow m15b6 has very similar structure
to the nonrotating s11 and s15 models. In the 15\mo\ E15A model, 
rotational effects lead to a considerably more extended iron core 
with lower central density and shallower density stratification than 
in the nonrotating s15. Model E20A's structure, however, 
does not deviate significantly from that of the nonrotating s20.
 
The models are mapped onto our computational grids according
to the same prescription presented for the first set: 
``shellular rotation'' in 1D, ``cylindrical rotation''
for most two-dimensional models. In naming the models belonging to this
set, we follow the conventions mapped out for the 
first set, but leave out the parts concerning parameters A and P: 
E15A:2Ds, for example, is progenitor E15A of \cite{heger:00} 
evolved in axisymmetry, but with ``shellular rotation.'' 
Table \ref{table:initialmodels} lists the initial model parameters 
for E15A, E20A, and m15b6.

\section{Results}
\label{section:results}
The focus of this study is on the periods and rotational profiles of 
the PNSs as obtained from our one-dimensional and 
two-dimensional radiation-hydrodynamics simulations. For more in-depth 
discussions of the supernova physics, we refer the reader to 
\cite{thompvisc:05} for the one-dimensional simulations with SESAME 
and to \cite{walder:05} for the two-dimensional MGFLD VULCAN/2D 
simulations. 

In the discussion of our one-dimensional simulations we limit
ourselves for conciseness to those performed with the s11 model (11~\mo\ at ZAMS) of
\cite{ww:95} while we describe and compare in detail simulations with
a wide variety of progenitor star models in the discussion of our
two-dimensional calculations. 

\subsection{One-dimensional Simulations}
\label{section:results1D}

An interesting consequence of the self-similar, quasi-homologous nature 
of the inner core's collapse 
is the preservation of 
its initial rotational profile in the PNS from
the precollapse state. In other words,
if the inner core is initially rotating rigidly (which is 
likely to be the case; see \citealt{heger:00,heger:04} and
Fig.~\ref{fig:initialrot}), then the inner PNS 
will be in solid-body rotation as well. 
There is, however, a region in any PNS that is rotating 
differentially ---
independent of the angular momentum distribution in the progenitor core.
Due to angular momentum conservation the core spins up during collapse.
Since the inner parts of the precollapse iron core collapse to 
smaller radii than the supersonically collapsing 
outer parts, differential rotation is a natural consequence of collapse.
However, solid-body rotation is
the lowest-energy state of a rotating body and viscous processes can  
drive the differentially rotating PNS into rigid rotation 
as it evolves, cools and condenses to its final configuration. 
As \cite{thompvisc:05} have pointed out, 
early and efficient postbounce dissipation of the
 shear energy stored in the differential rotation of the
PNS by viscous processes (e.g., the MRI) can significantly
affect the dynamics and might result in a successful explosion.
To estimate the rotation periods our PNSs
would have if they were in solid-body rotation, 
we introduce a moment of inertia-weighted 
mean angular velocity: 
\begin{equation}
\label{eq:omegaav}
\bar{\Omega}(\zeta) =
\frac{ \int_{r=0}^{R} \zeta \rho({\mathbf{r}}) \Omega({\mathbf{r}}) 
r^2_\perp \mathrm{d}^3 r}
{\int_{r=0}^{R} \zeta \rho({\mathbf{r}}) r^2_\perp \mathrm{d}^3 r}\, ,
\end{equation}
where $r_\perp$ is the distance from the rotation axis at any given
$\mathbf{r}$, $r = |{\mathbf{r}}|$ is the spherical radius, $R$ is
the outer radius of the computational domain and  $\rho(r)$ is the
matter density. In the one-dimensional case, we use the 
spherical radius for $r_\perp$.
We now define two mean rotation periods. The 
definition of the function $\zeta$ depends on the individual
mean being considered. The first mean period we define is
\begin{eqnarray}
\label{eq:pavrho}
\bar{\mathrm{P}}_{\rho 12} &=& 2\pi (\bar{\Omega}(\zeta))^{-1}\,\, , 
\mathrm{where}\nonumber \\
\zeta({\mathbf{r}}) &=& \left\{ \begin{array}{l}
                           1 \hspace*{0.2cm}\mathrm{if}
\,\,\rho({\mathbf{r}}) \ge 10^{12}\, \mathrm{g}\,\,\mathrm{cm}^{-3}\\
			   0 \hspace*{0.2cm}\mathrm{otherwise}
                           \end{array} \right. \,\,.
\end{eqnarray}
$\bar{\mathrm{P}}_{\rho 12}$ is the period that all material
with $\rho \ge \mathrm{10}^{12}$~g~cm$^{-3}$ would have if it were 
in solid-body rotation. We use this quantity to describe
the postbounce spin evolution of the PNS. It is
particularly useful because it captures the effects of accretion and 
contraction to higher densities and is meaningful in both 
one and two spatial dimensions. We adopt the density cut of 
10$^{12}$~g~cm$^{-3}$, roughly marking the characteristic 
density at which neutrinos decouple from the fluid,
to define the ``edge'' of the PNS 
for analysis purposes. Clearly, a PNS does not have
a sharp boundary. However, most of its material has 
densities above the assumed density cut.
The second mean period is
\begin{eqnarray}
\label{eq:pavM}
\bar{\mathrm{P}}_{\mathrm{M}} &=& 2\pi (\bar{\Omega}(\zeta))^{-1}\,\, ,
\mathrm{where}\nonumber \\
\zeta(r,\mathrm{M}) &=& \left\{ \begin{array}{l}
1 \hspace*{0.2cm}\mathrm{if}\,\,\mathrm{M}(r) \le \mathrm{M}\\
0 \hspace*{0.2cm}\mathrm{otherwise}  \end{array} \right.\,\,.
\end{eqnarray}
$\bar{\mathrm{P}}_{\rm M}$ is the period that the material interior to a 
given mass coordinate M would  have if it were in solid-body rotation. 
An unambiguous definition of this quantity is only possible in one 
spatial dimension, because, in two dimensions one cannot assume that the 
matter distribution remains spherically symmetric --- rotation and 
postbounce convection break spherical symmetry in any two-dimensional
model.

\begin{figure}
\begin{center}
\includegraphics[width=7.5cm]{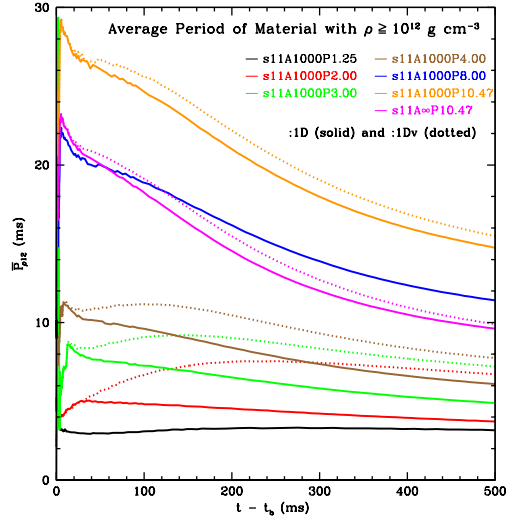} 
\caption{
Evolution of the moment of inertia-weighted 
mean period $\bar{\mathrm{P}}_{\rho \mathrm{12}}$ 
for material of density greater than or equal to 10$^{12}$ g cm$^{-3}$ 
according to eqs.~(\ref{eq:pavrho}) and~(\ref{eq:omegaav}). Shown
are selected models of the one-dimensional s11 model series with 
(dotted lines, :1Dv) and without (solid lines, :1D) viscous 
dissipation/angular momentum redistribution. For models 
s11A1000P1.25:1D and s11A1000P8.00:1D,
no computation with active viscous dissipation is performed.
Viscosity (see \S\ref{section:1Dmethods}) generically leads 
to spindown of the protoneutron
star. It is most efficient at doing so in fast rotators and
in the most extreme
case (in model s11A1000P2.00) leads to a final mean period that is 
twice as large as in the nonviscous case.
\label{fig:s11period}}
\end{center}
\end{figure}

The evolution of the postbounce mean
period $\bar{\mathrm{P}}_{\rho\mathrm{12}}$ as defined above is shown 
in Fig.~\ref{fig:s11period} for a variety of models
of the s11 model series. Shown in solid lines is the mean period for
 models without active viscous dissipation (:1D models). In dotted 
lines  we plot the corresponding models in which we include viscous 
dissipation (:1Dv models). Let us first focus on the evolution of 
\pav for models without the inclusion of viscous effects. During 
collapse, \pav remains undefined until the density
reaches the threshold value of \rtwelve. Initially it increases as more 
slowly rotating material achieves a density of $10^{12}$~g~cm$^{-3}$ 
and reaches its maximum just $\sim$1~ms before core bounce, when the 
lion's share of the inner core has crossed the density threshold, 
but has still not spun up to its bounce angular velocity. At bounce, 
\pav experiences a local minimum and then increases again as the core 
re-expands to a quasi-equilibrium configuration. 

Model s11A1000P1.25:1D (shown in black in 
Fig.~\ref{fig:s11period}) is our fastest rotator, with an
initial central period of 1.25~s and $\beta$ = T/$|W|$ $\sim$0.9\%.
It undergoes a core bounce altered by centrifugal forces and forms 
a quickly spinning PNS with a central period of 
$\sim$1~ms. As displayed in Fig.~\ref{fig:s11period}, its mean period 
\pav stays almost constant at $\sim$3 ms until the end of the 
calculation. Because of strong centrifugal support, its protoneutron 
star is large and, even
though matter is accreting, is only slowly contracting. 
Furthermore, model s11A1000P1.25:1D is the only one-dimensional 
model without viscous dissipation that exhibits an explosion 
--- though a weak one --- and less matter is accreted onto 
the PNS than in models that do not 
explode. Model s11A1000P2.00:1D 
is the next slowest rotator in our s11 model series. 
Its core bounce is dominated by nuclear repulsive forces, 
but centrifugal forces still play an important role. Its \pav increases 
slightly after bounce, reaches a local maximum at a time of 
$\sim$30~ms after bounce, and decreases from thereon to a 
final\footnote{In the following --- unless otherwise stated ---
``final'' means ``at the end of the numerical evolution.''}
 \pav of~3.5~ms.

With increasing initial central rotation period, the influence of 
centrifugal forces becomes less and less relevant. As can be inferred 
from Table \ref{table:results1D}, the initial value of $\beta$ 
is already as low as 0.25\% for model s11A1000P3.00. In 
Fig.~\ref{fig:s11period} all models with initial central period above 
3~seconds show the same qualitative behavior: a short increase of \pav 
after bounce and then a monotonic decrease as more material
is crossing the $10^{12}$ g cm$^{-3}$ density threshold and piling up 
on the PNS. The slowest model evolved without viscous 
dissipation, s11A1000P10.47:1D, has a final ($\sim$500 ms after bounce) 
mean period of $\sim$14 ms. Model s11A$\infty$P10.47:1D 
(magenta graph in Fig.~\ref{fig:s11period}) has a rotation 
parameter A (eq.~[\ref{eq:rotlaw}]) set to 
infinity, forcing the initial model to rotate like a solid body. 
Compared with 
model s11A1000P10.47:1D, s11A$\infty$P10.47:1D has significantly
more angular momentum in the outer regions of the core. This 
difference has a major impact on the PNS's rotation and 
leads to a $\sim$50\% smaller final mean period.  

Dissipation of the shear energy stored in differential rotation is, as 
\cite{thompvisc:05} report, most efficient in fast rotating models. 
This is supported by the evolution of our s11 models that include the 
effects of viscous energy dissipation and angular momentum 
redistribution  after bounce.  In Fig.~\ref{fig:s11period}, we plot in dotted lines the 
evolution of \pav for our :1Dv models. For model s11A1000P2.00, 
the :1Dv variant begins to deviate significantly from the model without 
viscous dissipation at about 40~ms after bounce. While the :1D model's 
mean period slightly decreases,  \pav of the :1Dv model increases 
strongly as long as the angular momentum redistribution mechanism 
provided by viscosity remains efficient, that is, as long as there is 
considerable shear energy in differential rotation that can be 
dissipated and as long as the viscous timescale is short with
respect to the Kelvin-Helmholtz timescale. 
Eventually, at $\sim$300 ms after bounce the entire
PNS is in solid-body rotation and \pav remains constant, or
decreases slightly, as the PNS contracts.
All fast models for which we have performed :1Dv evolution 
(s11A1000P2.00, s11A1000P2.34, s11A1000P3.00, s11A1000P4.00) experience
successful explosions 
(for details on the explosion mechanism, see \citealt{thompvisc:05}).
Since rotational energy scales with $\Omega^2$, there is, for the same
qualitative rotational profile, much less energy stored in the
differential rotation of slow models than in fast models. 
This is why slow one-dimensional rotators (P$_0$~\sgreat~5~s) do 
not explode
even with the additional heating provided by viscous dissipation. 
Furthermore, this comports with the fact that the change in \pav 
brought about by the inclusion of viscous effects operating on 
differential rotation 
becomes smaller with increasing initial period.
(Recall that the viscous timescale scales with $\Omega^{-1}$;
\citealt{thompvisc:05}.)

The global ratio of rotational kinetic energy T to 
gravitational potential energy $|$W$|$, 
the rotation parameter $\beta = \mathrm{T}/|\mathrm{W}|$, 
is another useful and important quantity whose evolution
we follow for all our models. If the value of $\beta$
passes some critical threshold value, 
nonaxisymmetric rotational fluid instabilities may set in and lead
to large-scale deformation of the core. In the approximation of
MacLaurin spheroids (uniform density, incompressible, rigidly
rotating equilibrium configurations), triaxial instabilities
may grow if $\beta$~$\ge$~14\% or $\beta$~$\ge$~27\% for secular and
dynamical instabilities, respectively (\citealt{tassoul:78}). Given
recent results by \cite{rotinst:05} (see also 
\citealt{cent:01,shibata:02a,saijo:03,shibata:04a}), it is likely
that the above canonical threshold values for $\beta$ do not hold
for differentially rotating postbounce cores with a realistic
equation of state. \cite{rotinst:05} have found that a 
dynamical m = 1--dominated spiral instability may set in at a 
$\beta$ as low as 8\%. Nonaxisymmetric
instabilities are of particular interest because they may lead to the
 emission of strong gravitational waves and to angular momentum
redistribution and spindown of the central core.

\begin{figure}
\begin{center}
\includegraphics[width=7.5cm]{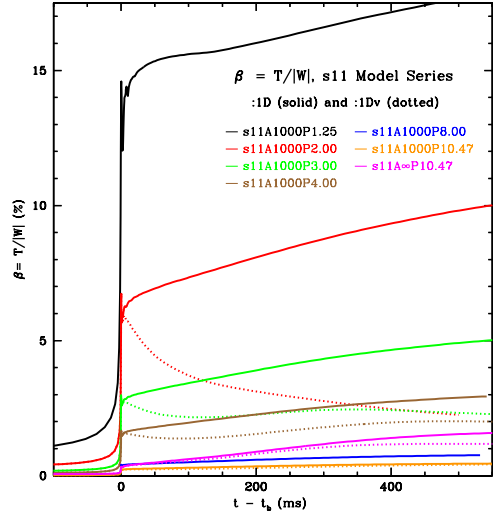} 
\caption{Evolution of the rotation parameter $\beta$ = T/$|$W$|$ for
selected models of the s11 model series in 1D. Solid lines: models
without viscous dissipation (:1D), 
dotted lines: models with viscous dissipation (:1Dv).
For models s11A1000P1.25:1D and s11A1000P8.00:1D,
no computation with active viscous dissipation is performed.
Note the strong impact of viscous dissipation acting on the
energy stored in differential rotation.
\label{fig:s11beta}}
\end{center}
\end{figure}

Figure \ref{fig:s11beta} shows the evolution of $\beta$
for selected :1D and :1Dv models of the s11 model series. 
s11A1000P1.25:1D is the fastest rotator in this series. Initially, 
$\beta$ is 0.88\% for this model and it reaches a local maximum of 14.6\%
at bounce. After bounce, the relaxing core, abetted by centrifugal 
forces, 
overshoots its equilibrium configuration, leading to a pronounced local 
minimum in $\beta$ of 12\% about 3~ms after bounce  
\footnote{Due to the range in time we are showing 
in Fig.~\ref{fig:s11beta},
this local minimum is not clearly visible.}. Immediately following the
initial postbounce relaxation, $\beta$ begins to increase and does
so throughout the rest of the evolution, reaching a $\beta_f$ 
(the $f$ indicating ``final'') of 18.4\% at the end of our numerical 
evolution. Given the approximate constancy of the mean period \pav 
of model s11A1000P1.25:1D (Fig.~\ref{fig:s11period}) the above 
might seem surprising. However, $\beta$ is a 
global quantity; the integrals for T and W encompass the entire 
grid and not just material above a density cut. Hence, the linear
(in time) increase in global $\beta$ indicates that the protoneutron 
star is slowly beginning to contract and spin up, but also  that more 
and more material from larger radii is falling in, spinning up and 
settling onto the PNS surface. Both T and $|$W$|$ 
increase, with T dominating, because of its higher-power dependence on 
the radial coordinate. 

The $\beta$ evolution for the rest of the s11 models without viscous
dissipation is essentially the same as the one just described for model
s11A1000P1.25:1D. With decreasing initial $\beta$, 
$\beta$ at bounce and $\beta_f$ decrease. Model
s11A$\infty$P10.47:1D, whose core is initially in solid-body rotation,
 stands out. Even though its initial period P$_0$ is the same as model
s11A1000P10.47:1D's, its initial value of $\beta$ is more than four times
larger (see Table~\ref{table:results1D}) than s11A1000P10.47:1D's. 
As shown in Fig.~\ref{fig:s11beta}, this
is reflected in its $\beta$ evolution up to bounce and, most apparently,
in the steeper increase in the postbounce phase.

Two of our models from the s11 model series, namely 
s11A1000P1.25:1D and s11A1000P2.00:1D surpass $\beta$ = 8\% during
their numerical evolution. They also exhibit strong differential 
rotation (Fig.~\ref{fig:s11final}) and might experience the
m=1-dominated one-armed spiral instability (\citealt{rotinst:05}).
s11A1000P1.25:1D even reaches $\beta$~$\ge$~14\% and may become 
unstable towards a ``classical'' bar-like instability on a secular 
timescale. 

\begin{figure}
\begin{center}
\includegraphics[width=7.5cm]{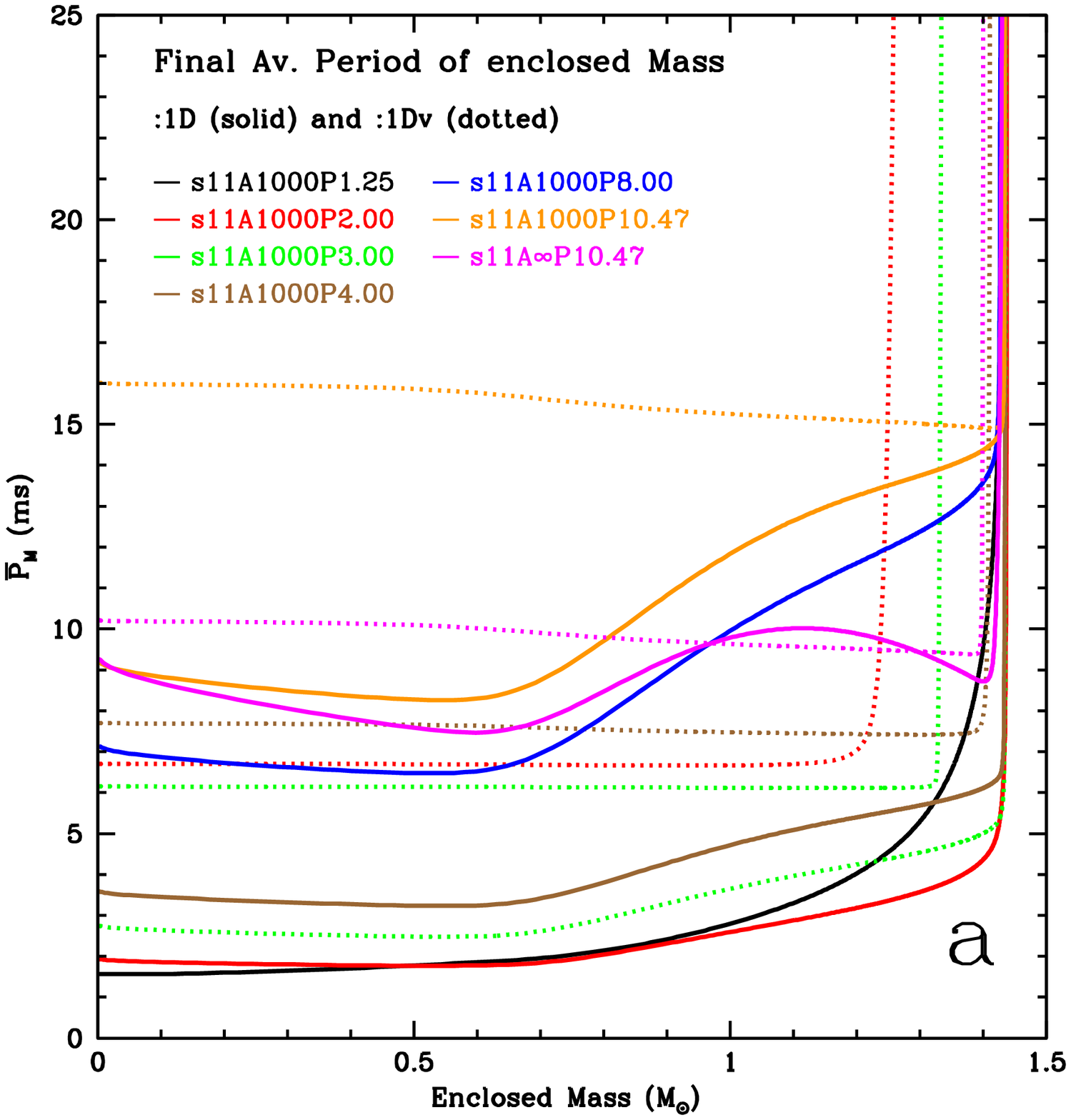} 
\includegraphics[width=7.5cm]{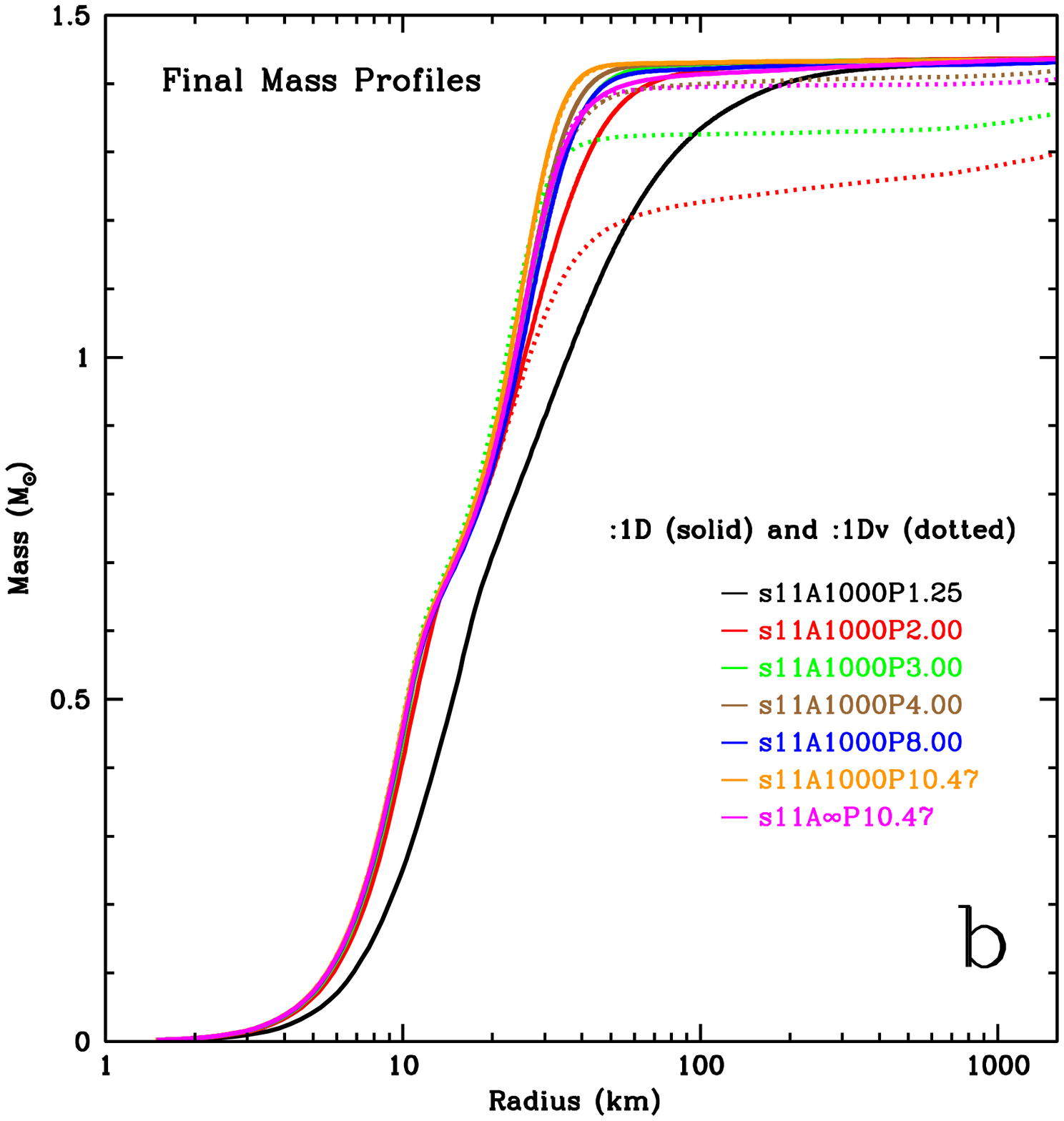} 
\caption{(a) Final moment-of-inertia-weighted mean period 
\pavm
for material interior to a given mass coordinate M and (b) mass
 M($R$) contained inside a radius $R$. Both quantities
are plotted for a subset of the s11:1D and :1Dv model series at the
end of each calculation. The early and strong explosions
of models s11A1000P2.00:1Dv and s11A1000P3.00:1Dv lead 
to significantly smaller PNS masses. This is
apparent in plot (b), but also in plot (a), where the lower
``mass cut" in \pavm is obvious. Models
s11A1000P1.25:1D and s11A$\infty$P10.47:1D, and their :1Dv variants, also explode,
but more weakly and at later times. This is why their profiles
do not deviate as much from the models that do not explode. 
Model s11A1000P1.25:1D is the fastest model that we consider.
Due to strong centrifugal forces, s11A1000P1.25:1D's protoneutron
star is rather extended (plot b). 
Also note that the inclusion of dissipative angular momentum 
redistribution leads to rigid rotation throughout the
PNS for all :1Dv models.
\label{fig:s11mass}}
\end{center}
\end{figure}

The evolution of $\beta$ for models with
viscous angular momentum redistribution is quite different
(Fig.~\ref{fig:s11beta}).
As soon as the viscous terms are switched on after bounce 
(\S\ref{section:1Dmethods}),  
for :1Dv models with P$_0$~\sless~4~s (red, green
and brown dotted lines in Fig.~\ref{fig:s11beta}) $\beta$ begins to
decrease. Shear energy stored in the differential rotation of 
the PNS is dissipated away, leading to
 solid-body rotation at lower angular velocities and, hence, 
lower rotational kinetic
energies. As discussed by \cite{thompvisc:05}, models
s11A1000P2.00:1Dv and s11A1000P3.00:1Dv experience energetic
explosions that remove a significant fraction of the stellar
envelope. Therefore the PNS masses remain somewhat 
smaller than for s11A1000P4.00:1Dv, which explodes less energetically
and at later times, allowing more matter to accrete through the 
stalled supernova shock and onto the PNS 
(Fig.~\ref{fig:s11mass}). In model 
s11A1000P4.00:1Dv, $\beta$ decreases for the first 100 ms and then 
increases again. This is, in part, because of the $\sim\Omega^3$ 
dependence of the viscous dissipation and, in part, because of 
its weak and later explosion that results in more mass 
being accreted than 
in the two faster models discussed. For even slower models,
the evolution of $\beta$ is similar to those for the cases 
without viscosity. Model
s11A$\infty$P10.47:1Dv departs from the :1D evolution at late times when
high-angular momentum material reaches small radii and the energy
stored in differential rotation becomes large. Viscous
dissipation of that energy then leads to a smaller final $\beta$ than
in the :1D model without viscous dissipation.

In Fig.~\ref{fig:s11final}, we present the final
period (Fig.~\ref{fig:s11final}a) and angular velocity profiles
(Fig.~\ref{fig:s11final}b) of a representative subset of the s11 model
series calculated with SESAME. Profiles of models with viscous 
dissipation and angular momentum redistribution are drawn as dotted 
lines, profiles of models without viscous effects are drawn as 
solid lines. 
Model s11A1000P1.25:1D, drawn in black, has the shortest initial 
central period (P$_0$~=~1.25~s) which maps to the shortest final central 
period of P$_{0,f}$~=~1.56~ms. As discussed earlier in this section, 
the PNSs are in solid-body rotation inside  8 -- 10~km 
(or $\sim$0.5~--~0.8~\mo\ in mass coordinate). At greater radii, the 
PNSs quickly become differentially rotating. For model 
s11A1000P1.25:1D, the  angular velocity drops from 
$\sim$2500~rad~s$^{-1}$ at 20~km to $\sim$900~rad~s$^{-1}$ at 50~km 
and $\sim$330~rad~s$^{-1}$ at 100~km, corresponding to a 
$\Delta \Omega / \Delta r$ of 53~rad~s$^{-1}$~km$^{-1}$
from 20 to 50~km and 11~rad~s$^{-1}$~km$^{-1}$ from 50 to 100~km 
(be aware of the log-log scale in Fig.~\ref{fig:s11final}). 
Again, one has to keep in mind that model s11A1000P1.25:1D explodes. 
The explosion removes material that would otherwise accrete onto the 
PNS and pile up on the quasi-incompressible inner region 
above radii of 15 -- 30~km. This is
reflected in the features visible in the graphs of :1D models with
P$_0$~$\ge$~2 s in the range of radii mentioned. Models in which we
include viscous dissipation and angular momentum transport exhibit
solid-body rotation out to much larger radii. Model s11A1000P2.00:1Dv,
for example, rotates rigidly out to $\sim$23~km or 0.9~\mo\ in mass
coordinate while model s11A1000P2.00:1D exhibits solid-body rotation
out to only $\sim$10~km or 0.42~\mo \footnote{See Fig.~\ref{fig:s11mass}b for 
the mapping between mass and radial coordinates at the end of each
model's numerical evolution.}. Viscous dissipation 
of shear energy stored in differential rotation also leads to 
a spindown of the PNS. s11A1000P2.00:1D's final
central period is 1.95 ms whereas s11A1000P2.00:1Dv spins with
a final central period of 6.7 ms, which is more than 3.5 times slower.
Due to the $\Omega^2$ dependence of rotational energy and the
$\Omega^3$ dependence of the dissipation rate, the relative 
period increase brought about by viscous dissipation/angular 
momentum transport decreases quickly with increasing P$_0$. For
s11A1000P3.00 (central periods: 2.74 ms and 6.16 ms for the 
:1D and :1Dv models, respectively), the ratio is down to about 2.2 and 
for s11A1000P10.47 (central periods of 9.19 ms and 16.0 ms for 
models :1D and :1Dv, respectively), it is approximately 1.7. Of course, 
the just quoted ratios depend on the simulation time after bounce, which
determines how much time we have allowed the dissipative process to spin 
down the core and how much material and angular momentum is accreted 
during the simulation. Model s11A$\infty$P10.47 stands out. It contains 
significantly more angular momentum than its counterpart with 
A~=~1000 km. Therefore, it spins up to shorter central periods during 
collapse and during the postbounce evolution. 

For key models of the
s11 model series, the final moment-of-inertia-weighted mean period \pavm,
as defined by eq.~(\ref{eq:pavM}), is shown in Fig.~\ref{fig:s11mass}a. 
Note that \pavm gives the mean period of material 
inside a mass coordinate M, assuming solid-body rotation.
Hence, the graphs should not be interpreted as profiles in the
usual sense. The mass coordinate  
at which the slope of each individual model's \pavm 
sharply steepens is a good estimate for the 
compact remnant mass in models that exhibit explosions. 
s11A1000P2.00:1Dv (dotted red), 
for example, explodes early and energetically,
leaving behind a lower-mass PNS than model
s11A1000P3.00:1Dv (dotted green), 
which explodes at a somewhat later time. For models
that don't explode, \pavm jumps only at $\sim$1.42~\mo. 
The consequences of viscous dissipation and
angular momentum redistribution in the :1Dv models, 
spindown and solid-body rotation, are perhaps even more
obvious in Fig.~\ref{fig:s11mass}a than in Fig.~\ref{fig:s11final}.
Also, note that, since the :1Dv models are in solid-body
rotation throughout most of the protoneutron
star at the end of our calculations, \pavm for them
agrees very well with the actual period profiles. 

\newpage

\subsection{2D Simulations}
\label{section:results2D}

\begin{figure}
\begin{center}
\includegraphics[width=8cm]{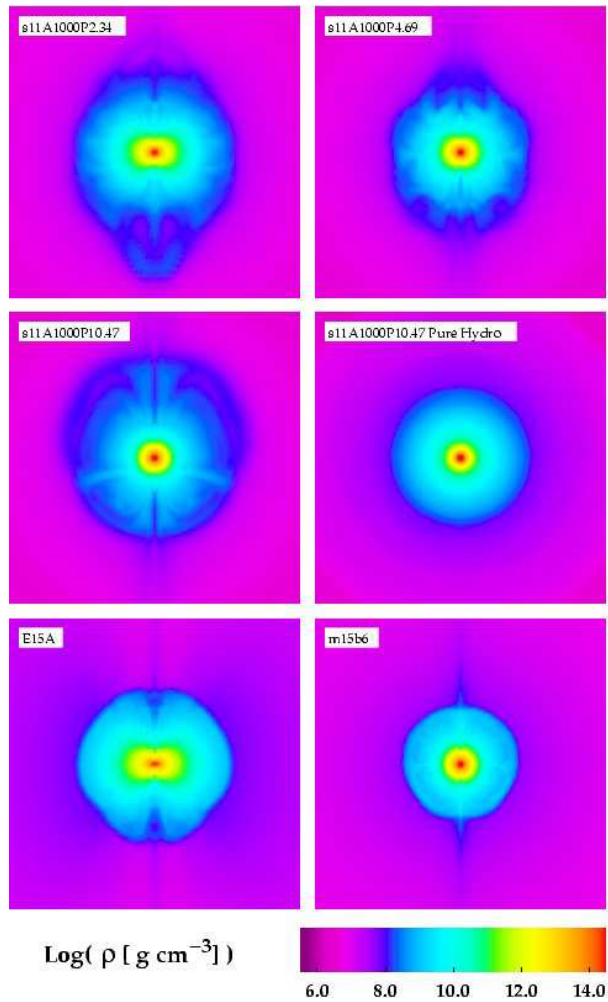} 
\caption{Logarithmic color map of the density 
for a variety of 2D models at 200 ms after core bounce.
Shown is the inner 400 km on a side. This figure nicely
demonstrates the pronounced rotational flattening of the protoneutron
star in fast models (s11A1000P2.34 and E15A). With increasing
precollapse iron core period, the PNS
becomes less and less rotationally flattened. In models
s11A1000P10.47 and m15b6 the flattening has almost disappeared.
The lighter cores (s11 models) show considerably stronger
convective structure --- with the exception of the purely
hydrodynamic variant of s11A1000P10.47 --- than the more
massive cores (E15, m15b6) at 200 ms after bounce. The 
slight ``dents'' and ``spikes'' along the rotation axis are
artefacts of the imperfect symmetry axis treatment.
Note that the pure hydro variant of s11A1000P10.47 was 
evolved with an updated version of the VULCAN/2D with 
improved axis treatment.
\label{fig:panel2Ddensity}}
\end{center}
\end{figure}

We now discuss the results of the axisymmetric 2D VULCAN simulations 
using the methods described in \S\ref{section:2Dmethods}.
In contrast to the spherically symmetric simulations,  
two-dimensional, azimuthally-symmetric simulations are able to 
follow the non-spherical effects of rotation in core collapse. 
The degree of oblate deformation can be described 
by the ratio of the PNS's polar and equatorial 
radii on isodensity surfaces, the so-called axis ratio. 
Figure~\ref{fig:panel2Ddensity} 
displays snapshots of the density at $\sim$200 ms after bounce for 
a set of representative 2D models. Models s11A1000P2.34 and E15A are 
the fastest rotators, having axis ratios below 1:2. Clearly, 
a two-dimensional treatment is a must for capturing the
full dynamics of these rapid rotators.
With increasing initial iron core period (P$_0$) the PNS
axis ratio increases towards one. m15b6 and the two 
variants of s11A1000P10.47:2D exhibit almost no rotational 
flattening at all and one might expect a one-dimensional treatment to
yield good results. But it is not only rotation that is modeled much
more realistically in real 2D simulations. Neutrino-driven convection 
naturally develops in the region behind the stalled 
supernova shock and leads to the interesting substructure 
seen in Fig.~\ref{fig:panel2Ddensity}. 
The purely hydrodynamic model, s11A1000P10.47:2Dh, experiences
a prompt explosion and Fig.~\ref{fig:panel2Ddensity} does not show the
imprint of the vigorous convection present in the models
that are evolved with adequate neutrino treatment and radiative
transfer. Note that the small dents and spikes visible 
along the rotation axis in most of the 
panels are related to the imperfect symmetry axis treatment
in the version of VULCAN/2D with which most of our models
are evolved. The purely hydrodynamic model has been evolved with an
updated version of the code and exhibits minimal axis artefacts.

\begin{figure}
\begin{center}
\includegraphics[width=7.5cm]{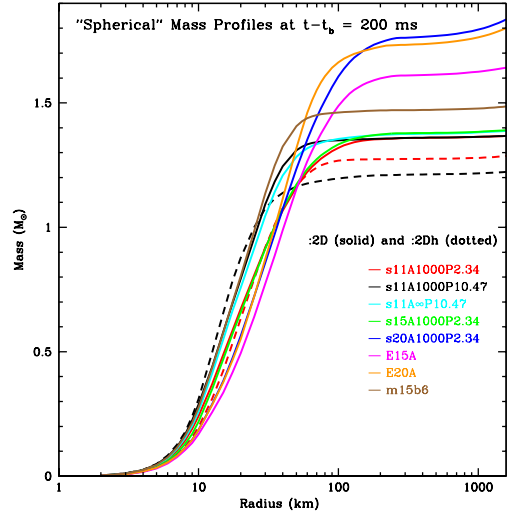} 
\caption{Mass contained inside a sphere of radius R for
selected two-dimensional models $\sim$200~ms after core bounce. 
The ``s" models (\citealt{ww:95}) with P$_0$ = 2.34 s and the fast
E15A and E20A models (see model parameter details in 
Table~\ref{table:initialmodels}) deviate significantly from
spherical symmetry. Their spherical mass profiles should, hence,
be taken with caution. Solid lines represent mass profiles 
of MGFLD models. Dashed lines correspond to models which 
are evolved purely hydrodynamically and experience prompt
explosions that blast away the outer core.
\label{fig:intermodel_2D_mass}}
\end{center}
\end{figure}

Because of the computational difficulties that
accompany long-term, two-dimensional radiation-hydrodynamics
evolutions, we have run our two-dimensional models to postbounce 
times of only 200~--~300~ms and display profiles 
and two-dimensional snapshots at 200 ms after bounce. At that epoch, the 
PNS has already reached a compact state and the largest
fraction of its final mass has already settled inside a 
(spherical) radius of $\sim$50 km. 
Figure~\ref{fig:intermodel_2D_mass} shows the mass contained
inside a given spherical radius for selected models at 200~ms after
core bounce. The graphs certainly ignore
the two-dimensional nature of the calculation, but give a good
(and for the slow models, almost accurate) 
handle on the mass-radius relationship. The dashed lines in
Fig.~\ref{fig:intermodel_2D_mass} represent mass profiles
of models that are evolved adiabatically 
and exhibit prompt explosions, blasting away a large fraction
of the outer iron core and the stellar envelope. Hence, their 
PNS masses are small compared with those of the 
MGFLD models that might (or might not) explode at later times
after much of the outer core has accreted onto the PNS.
Note that slower models have more compact PNSs
than faster models. This is, of course, expected since rotational
support is greater in faster models and leads to more
expanded postbounce configurations. 

\begin{figure}
\begin{center}
\includegraphics[width=7.5cm]{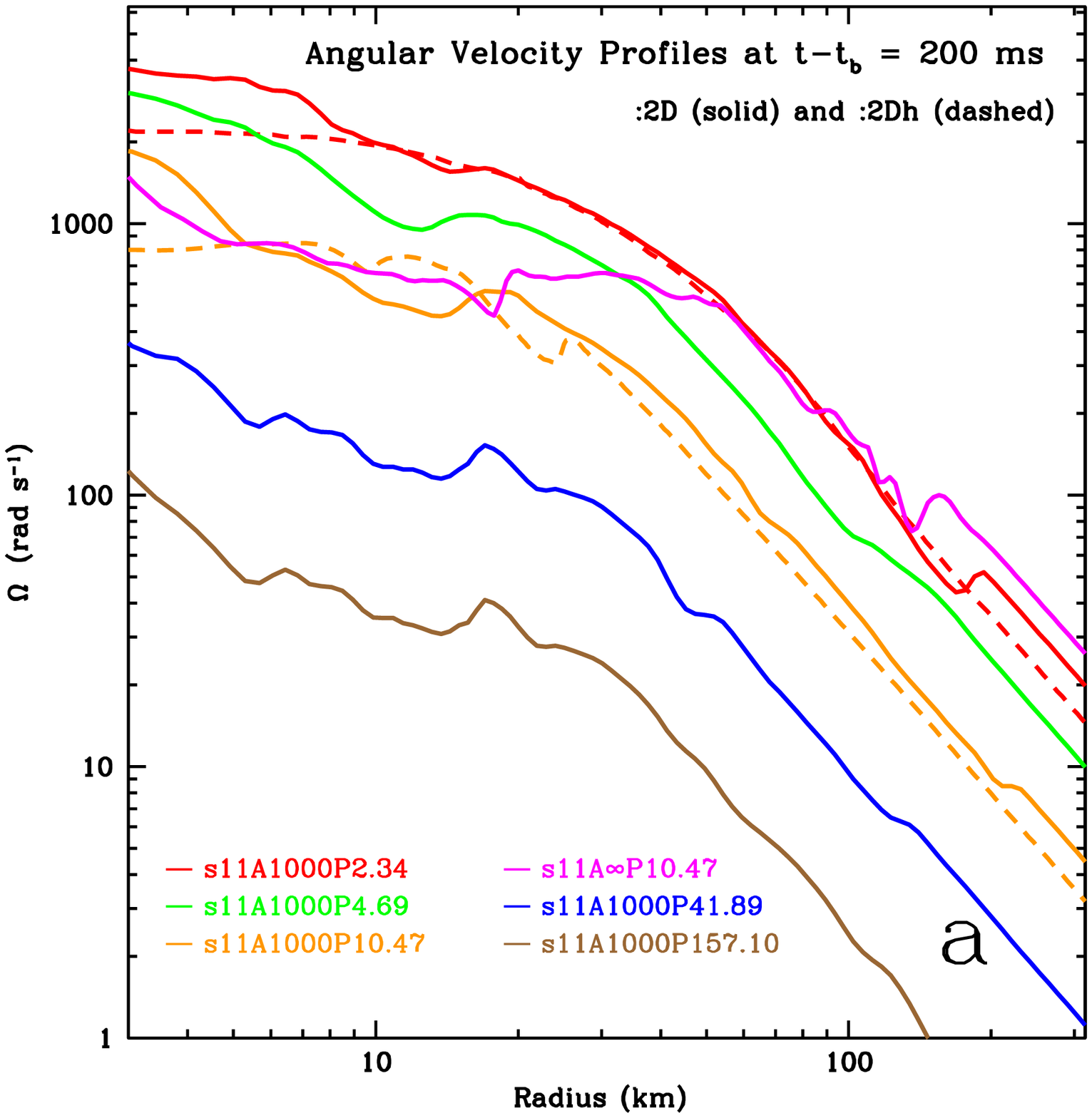}
\includegraphics[width=7.5cm]{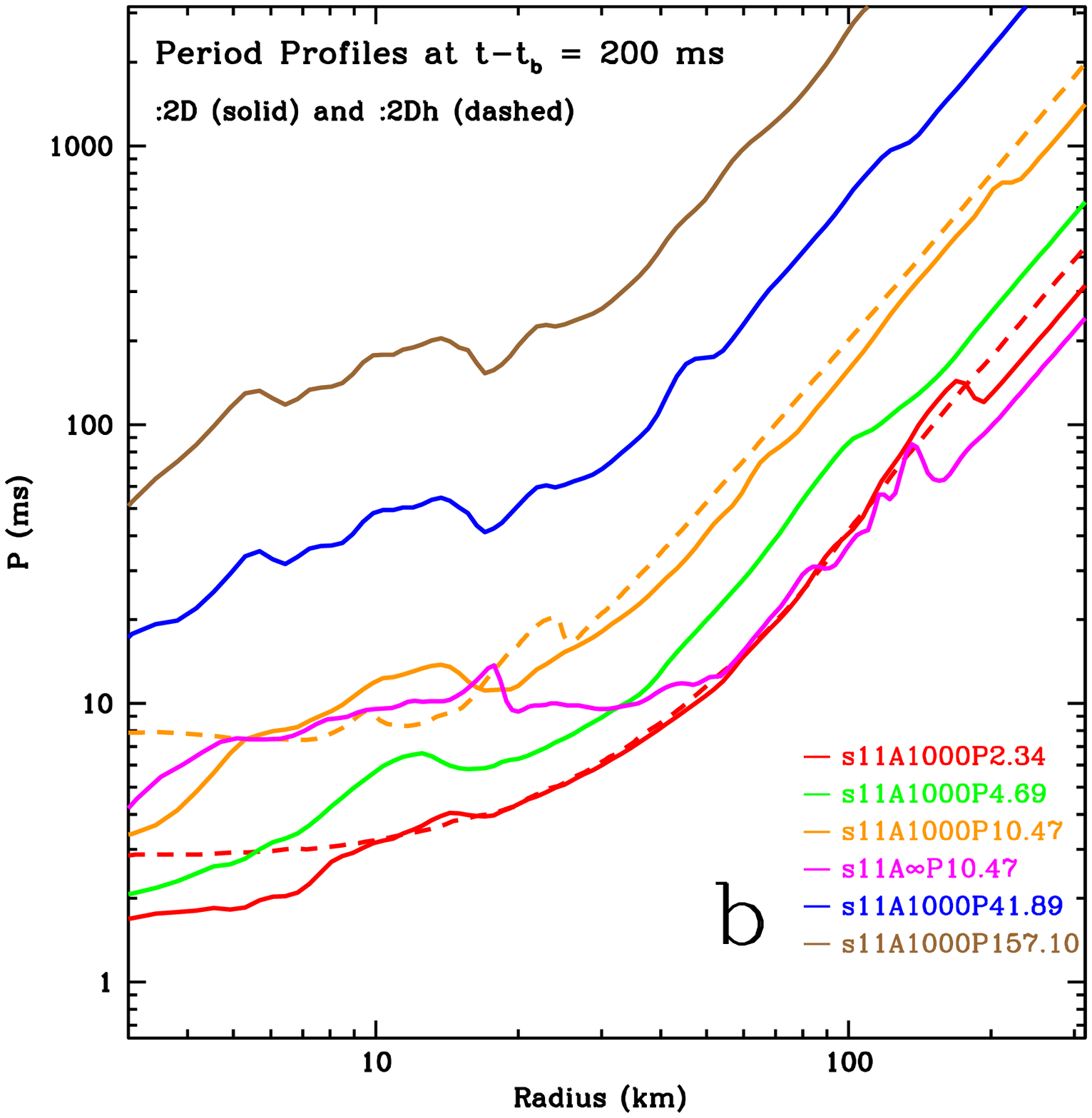}
\caption{Equatorial profiles of the angular velocity (a) and 
the period (b) at 200 ms after bounce for each model of our 
s11:2D model series.  The two-dimensional PNSs exhibit a 
low degree of differential rotation inside $\sim$20~km and then 
become strongly differentially rotating further out. All s11A1000:2D
models exhibit the same qualitative features. Model 
s11$\infty$P10.47 (initial solid-body rotation), however, 
differs from its A~=~1000~km counterpart.
Its PNS is appreciably less compact and the region of
quasi-solid-body rotation is more extended. The dashed lines correspond
to models that are evolved adiabatically. These models
promptly explode and do not undergo a phase of strong convection. Hence,
their equatorial profiles show less structure.
The profiles are plotted from 3 km on outwards. Due to the imperfect
axis treatment in the version of VULCAN/2D that is used for 
these calculations, we do not highlight the angular velocity and period 
data inside $\sim$3 km.
\label{fig:s11_2D_omegaperiod}}
\end{center}
\end{figure}

\begin{figure}
\begin{center}
\includegraphics[width=9cm]{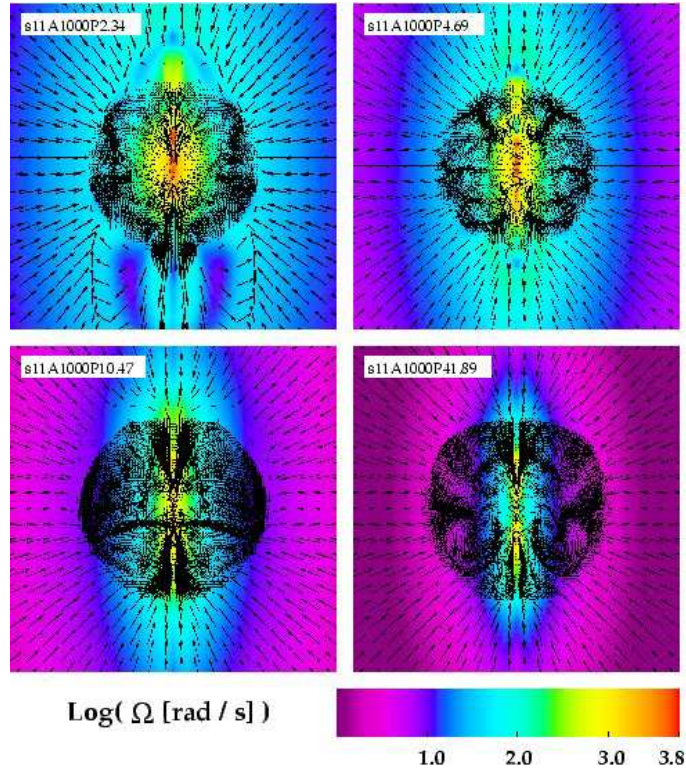} 
\caption{Logarithmic color map of the angular velocity $\Omega$
for models of our s11A1000:2D model series evolved 
in axisymmetry using VULCAN/2D 
with P$_0$~=~2.34~s, 4.69~s, 10.47~s, and 41.69~s, 
corresponding to $\Omega_0$~=~2.68~rad~s$^{-1}$,
1.34~rad~s$^{-1}$, 0.60~rad~s$^{-1}$, 0.15~rad~s$^{-1}$. 
Shown are the inner 400 km on a side. Fluid velocity vectors
are superposed. The vector data are downsampled at radii larger 
than $\sim$200~km. The panels show the
angular velocity distribution at about 200~ms after 
bounce in each model. The fastest model, s11A1000P2.34:2D, 
has a large centrifugally supported region of high angular-velocity 
material. This region grows increasingly smaller with increasing
initial iron core period. The color map has been artificially
limited to a maximum of 10$^{3.8}$ rad s$^{-1}$ to cut out
unphysically large angular velocities in the zones very close to the
rotation axis.\label{fig:s11_2Dpanel}}
\end{center}
\end{figure}

In Fig.~\ref{fig:s11_2D_omegaperiod}, we depict the equatorial
angular velocity and period profiles (panels (a) and (b), 
respectively) at 200~ms after bounce for the two-dimensional
s11A1000 model series. For comparison we also display
graphs of model s11A$\infty$P10.47:2D (magenta; initially in solid-body
rotation) and of two purely hydrodynamic models (dashed graphs).
Note that the graphs start at a radius of 3~km to excise
axis artefacts. The aim of Fig.~\ref{fig:s11_2D_omegaperiod}
is to display the systematic dependence of the protoneutron
star rotational profile on the initial iron core spin rate.
For this, we focus in Fig.~\ref{fig:s11_2D_omegaperiod} on the 
angular velocity
and period profiles of the MGFLD models 
s11A1000P157.10:2D, s11A1000P41.69:2D, s11A1000P10.47:2D,
s11A1000P4.69:2D, and s11A1000P2.34:2D. 
These models form a sequence in initial rotation period
from 157.10~s down to 2.34~s. At 200~ms after bounce, 
s11A1000P157.10:2D has reached a central period of~$\sim$50 ms. 
The spin period increases by about a factor of two  
from 3~km to 5~km and then stays roughly
constant, but shows significant substructure, out to 30~km.
From 30~km to 100~km, the period increases by a factor of ten,
corresponding to a ``period slope'' of~$\sim$30~ms~km$^{-1}$.
We call the reader's attention to the region
of quasi solid-body rotation between roughly 5 and 30~km 
also present in all spherically symmetric models 
(see \S\ref{section:results1D} 
and Fig.~\ref{fig:s11final}
). This region is a direct consequence of  
self-similar collapse of the inner core. The equatorial 
angular velocity and period profiles of the 
two-dimensional models are, however, much less smooth than
the profiles of the one-dimensional models 
. This is attributed to the
much more complicated, two-dimensional dynamics that involve
convection and inviscid angular momentum redistribution
along the rotation axis due to convection 
(Fig.~\ref{fig:s11_2Dpanel}). The hump in the
angular velocity profile (corresponding to a
notable dent in the period profile) 
at about 20 km is a generic feature in the entire s11A1000 series,
but is most pronounced in the slow models. It is caused
by accreting relatively high-specific angular momentum material
that is piling up on the nuclear-density region of the 
PNS. The faster a model is, the more extended
its PNS (see Fig.~\ref{fig:intermodel_2D_mass}) and
the shallower its density profile.

\begin{figure}
\centerline{
\includegraphics[width=7.5cm]{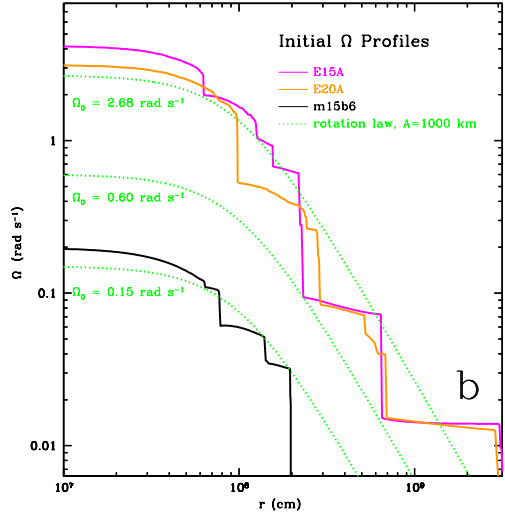} }
\caption{Initial angular velocity of the rotating progenitor models E15A (magenta), 
E20A (orange), and m15b6 (black) and of models that are set into rotation 
according to the rotation law of eq.~(\ref{eq:rotlaw}), with A fixed to 
1000 km (shown in green for various P$_0$ or $\Omega_0$). 
The discontinuities in the E15A, E20A, and m15b6 profiles 
are connected to compositional interfaces in the rotating progenitors
(see \citealt{heger:00}).
\label{fig:initialrot}
}
\end{figure}

The qualitative shape of the angular velocity and period profiles of 
s11A1000P157.10:2D is mirrored, but of course shifted, in 
the $\Omega${s} (and periods) in faster models 
(Fig.~\ref{fig:s11_2D_omegaperiod}). 
However, the profiles of faster models are smoother. Convection in 
fast models is largely suppressed by rotation in the equatorial 
regions \footnote{Note, however, that for convection to be stabilized,
the rotation rate must be large enough that if the calculation
had been done in magneto-hydrodynamics, the region would likely
be destabilized by the MRI (\citealt{thompvisc:05}).  We
therefore expect that when these rapidly rotating models
are calculated in MHD, they will show enhanced (magneto-)convection
with respect to the hydrodynamical, centrifugally-stabilized
calculations presented here.}
(\citealt{fryerheger:00}; \citealt{fryerwarren:04}; \citealt{walder:05}) 
and, hence, does not create as much substructure in the equatorial
profiles as in slower models. In Fig.~\ref{fig:s11_2D_omegaperiod}, 
for models s11A1000P2.34:2D and s11A1000P10.47:2D, we include equatorial
profiles of their purely hydrodynamic :2Dh variants. They are plotted 
with dashed lines. Since we do not include
any neutrino physics and transport in the :2Dh simulations,
these models cannot contract and spin up to higher $\Omega$ 
after core bounce. They experience prompt explosions and 
fall-back accretion is negligible. Keeping this in mind,
it is a bit surprising to see how similar the hydrodynamic
models' profiles are to the profiles of their MGFLD counterparts. 
To conclude our discussion of Fig.~\ref{fig:s11_2D_omegaperiod},
we call attention to the graphs of
s11A$\infty$P10.47:2D's angular velocity and period profile 
(magenta lines). This model is set up in solid-body rotation 
with an initial spin period of 10.47~s. Initial solid-body 
rotation means that a great part of the iron core's 
angular momentum is located at large radii (the $r^2$ 
moment arm balances the density drop with
increasing radius). Due to angular momentum conservation
during collapse, one expects the outer core layers to 
spin up appreciably. This is exactly what we see in 
Fig.~\ref{fig:s11_2D_omegaperiod}: The region of
high angular velocity ($\sim$600~rad~s$^{-1}$)
and low period ($\sim$10~ms) extends out to
about 70~km, whereas s11A1000P10.47:2D's  
angular velocity (orange lines) starts dropping (and its period
starts increasing) already at about 20~km. Note that the
centermost regions of the two models agree relatively well
--- as they should --- given the fact that s11A1000P10.47:2D's
initial angular velocity is almost constant throughout
its inner core (Fig.~\ref{fig:initialrot}).

The fastest model of the series presented in 
Fig.~\ref{fig:s11_2D_omegaperiod} is s11A1000P2.34:2D. Its 
initial central iron core spin (P$_0$) of 2.34~s maps to a central
postbounce period (at the 200~ms snapshot shown here) of about
1.7~ms, which increases to $\sim$3.2~ms at 10~km. As shown in
Fig.~\ref{fig:panel2Ddensity}, its PNS is
rotationally flattened with an axes ratio below 1:2. 
In Fig.~\ref{fig:s11_2Dpanel}, we show the angular velocity
distribution in s11A1000 models with P$_0$s of 2.34~s, 4.69~s,
10.47~s, and 41.89~s at 200 ms after bounce. 
Since we are imposing axisymmetry, the angular velocity distribution
is symmetric with respect to the vertical axis (the rotation axis).
We superpose flow velocity vectors to convey  
qualitatively the flow dynamics in the PNS. We
downsample the number of velocity vectors at radii that are 
greater than the approximate shock location of each 
individual model. For the purely hydrodynamic model, the 
radius beyond which we downsample is chosen to be 150~km. 
The panels show the inner 400~km of the computational domain.
(For a more detailed discussion of the multi-dimensional 
flow dynamics and its consequences concerning the supernova mechanism, 
we refer the reader to \citealt{walder:05}.) All 2D models 
shown rotate in the precollapse
stage with constant angular velocity on cylindrical shells. As
the collapse proceeds, the matter falls in almost spherically 
(more spherically for slowly spinning models, less spherically for
fast spinning models). The initial conditions and the
collapse dynamics are reflected in the elongated (prolate) shape
of the angular velocity distribution of the PNSs. The
fast models (s11A1000P2.34:2D and s11A1000P4.69:2D) have 
a notable central spheroidal bulge of high angular velocity material, 
which is formed by high specific angular momentum material from 
the equatorial regions.

\begin{figure}
\begin{center}
\includegraphics[width=7.5cm]{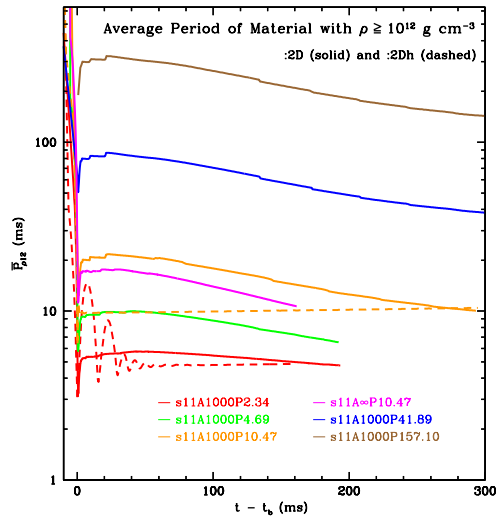} 
\caption{Postbounce evolution of the moment of inertia-weighted 
mean period \pav. Solid-line graphs depict standard models 
of the s11A1000:2D model series. Plotted with the dashed lines 
are the \pav of models that were evolved purely hydrodynamically. 
As expected, all \pav exhibit a local minimum at bounce when the 
inner core is most compact. \pav increases slightly after bounce, 
but ultimately decreases as more material settles onto the 
contracting PNS. With increasing initial
iron core spin period, \pav increases systematically. The purely 
hydrodynamic models' PNSs cannot contract. 
Hence, their \pav{s} stay 
constant or increase slightly as lower angular momentum material 
settles above the density threshold. The fast pure hydro 
model (P$_0$~=~2.34~s) exhibits strong postbounce
oscillations like those described in the study of \cite{ott:04}. 
Such oscillations are critically damped in simulations 
including adequate neutrino treatment.
\label{fig:s11_2D_logperiod}}
\end{center}
\end{figure}

As in the discussion of our one-dimensional models
in \S\ref{section:results1D},  
we define a mean, moment-of-inertia-weighted period
\pav for all material with density above 10$^{12}$~g~cm$^{-3}$
 (eqs.~\ref{eq:omegaav}~and~\ref{eq:pavrho}). 
Figure~\ref{fig:s11_2D_logperiod}
depicts \pav as a function of time for all two-dimensional
s11 models. 
In addition to the s11A1000 MGFLD models, we plot \pav for 
s11A$\infty$P10.47:2D (magenta) and the purely hydrodynamic variants
s11A1000P2.34:2Dh and s11A1000P10.47:2Dh (dashed lines). As one would
expect, all models exhibit a local minimum of \pav at the time of
core bounce. For the purely hydrodynamic models, this is the absolute
minimum in \pav, since they quickly achieve hydrostatic
equilibrium after bounce. Their periods remain almost constant
at postbounce times. Model s11A1000P2.34:2Dh (dashed red) undergoes core
bounce aided by centrifugal forces and shows the pronounced
postbounce coherent expansion and recontraction cycles described
for these sorts of models in \cite{ott:04}. In contrast, 
s11A1000P2.34:2D, its MGFLD variant, relaxes from its local 
minimum in \pav and shows no oscillatory behavior at 
all. We attribute this major difference to the following
phenomenon. In adiabatic collapse, 
angular momentum conservation during collapse leads to an
increase in the kinetic energy stored in 
rotation. By its nature rotation acts approximately like a 
gas with $\Gamma$ of 5/3 (\citealt{tassoul:78}; \citealt{mm:91}) 
and in this way effectively increases the pressure
support of the collapsing core.
At the moment of strongest compression at bounce,
centrifugal forces are maximal and result in the 
core dramatically overshooting its postbounce
equilibrium position. Recontraction follows, leading to another 
expansion-recontraction-bounce cycle. Eventually, 
enough pulsational energy is dissipated by the generation of 
secondary shocks that after several cycles the oscillation is damped.
However, including the production and radiative transfer 
of neutrinos, the rebounding core loses a large amount of energy 
in the neutrino burst that occurs when the bounce shock wave 
breaks through the electron neutrinosphere.
This, and continuing neutrino loses, lead to enhanced damping
of the postbounce expansion-recontraction-bounce cycles,
far more than experienced in 
quickly spinning purely hydrodynamic models. 

We note that the qualitative shape for a given progenitor structure
and rotational ``scale height'' A of the \pav 
evolution shown in Fig.~\ref{fig:s11_2D_logperiod} 
is the same for all s11A1000 models. Generally,  
at any postbounce time \pav 
is a monotonic function of the initial iron core spin. 
At 200~ms after core bounce, the largest
\pav is $\sim$180~ms (s11A1000P157.10:2D) and the smallest
is $\sim$4.7~ms (s11A1000P2.34:2D). 
In Table~\ref{table:results2D}, we list the mean periods 
for all models at the end of their numerical runs.
The magenta graph in Fig.~\ref{fig:s11_2D_logperiod} represents
s11A$\infty$P10.47:2D's \pav evolution. As expected,
this initially rigidly rotating
model's PNS is spinning faster than its counterpart with 
A set to 1000~km before collapse. 

\begin{figure}
\begin{center}
\includegraphics[width=7.5cm]{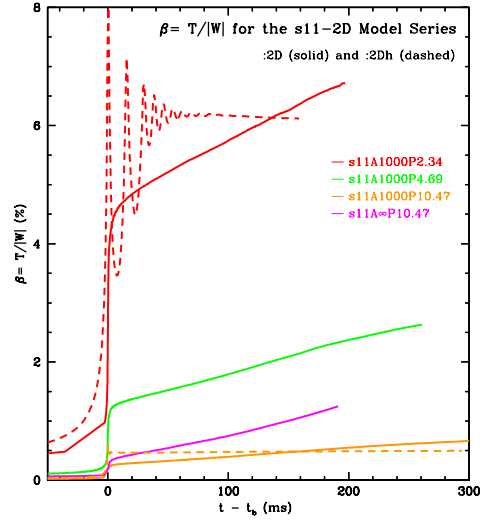} 
\caption{Evolution of the rotation parameter $\beta$ = T/$|$W$|$ 
at postbounce times for a subset of the s11A1000:2D model series.  
$\beta$ increases dramatically at bounce and increases monotonically, 
but linearly after bounce in all MGFLD
models, because of angular momentum conservation (and hence spin-up) 
of material that settles down onto the PNS. The slope 
of $\beta$ increases with decreasing precollapse spin period. Despite 
having the same initial spin (P$_0$~=~10.47~s), model 
s11A$\infty$P10.47:2D,
which is initially in solid-body rotation, reaches higher postbounce
$\beta$s than model s11A1000P10.47:2D. This is clearly related to the 
greater total angular momentum in the former. Model s11A1000P2.34-2Dh's 
$\beta$ evolution reflects this model's large-scale 
coherent postbounce oscillations. 
Both :2Dh models evolve to constant $\beta$, both because they 
explode and because their PNSs cannot contract and 
spin up.
\label{fig:s11_2D_beta}}
\end{center}
\end{figure}

As demonstrated for the
s11 model series in Fig.~\ref{fig:s11_2D_beta}, 
the systematics with the global rotation parameter $\beta$ 
is similar. Up to bounce, $\beta$ increases monotonically in all models. 
The faster of the two purely hydrodynamic models (dotted-line graphs) 
reaches the highest $\beta$ at bounce of all models presented in 
Fig.~\ref{fig:s11_2D_beta}. Due to its higher electron fraction $Y_e$
and higher effective adiabatic $\Gamma$ (\citealt{liebendoerfer:04}),
in direct contrast with the more realistic MGFLD variant,
s11A1000P2.34:2D, the purely hydrodynamic model has a much more
massive inner core.  Hence, during its ``plunge''
phase just before and at bounce, more material, and in that way
more angular momentum, reaches small radii. This is why
purely hydrodynamic models reach higher $\beta${s} at bounce than
their MGFLD diffusion counterparts. While the purely hydrodynamic
models settle at a practically constant final $\beta$, all 
MGFLD models exhibit a linear postbounce growth of the rotation 
parameter, with a slope that depends on the initial iron core spin and 
on the scale parameter A in the rotation law. The increase in
$\beta$ is caused by the increase in compactness and by
accretion of outer-core material. One might expect that the
slope of $\beta$ will change after a successful explosion (not
tracked in our models). At the end of their numerical evolution
none of the models presented in Fig.~\ref{fig:s11_2D_beta} has reached
the classical threshold values of $\beta$ for secular or dynamical 
instability. However, as \cite{rotinst:05} have demonstrated,
at $\beta$ $\sim$8\% realistic postbounce cores might undergo the 
low-T/$|$W$|$ instability. This value is likely to be reached by model 
s11A1000P2.34:2D at late times.

Having described the quantitative and qualitative features
of a neutron star's birth spin and its dependence on 
the initial iron core spin rate for a fixed
progenitor structure, we now investigate the effect
of varying progenitors on the PNS
spin. We have performed a set of calculations  
employing the s15, s20 (15~\mo\ and 20~\mo\ at
ZAMS, respectively) progenitors of \cite{ww:95}, the
E15A and E20A presupernova models of \cite{heger:00}, and 
the m15b6 progenitor of \cite{heger:04}. The key
progenitor characteristics of all models are
summarized in Table \ref{table:initialmodels}.
In the following we present the results of these calculations 
and compare them with those of our s11 models.

\begin{figure}
\begin{center}
\includegraphics[width=9cm]{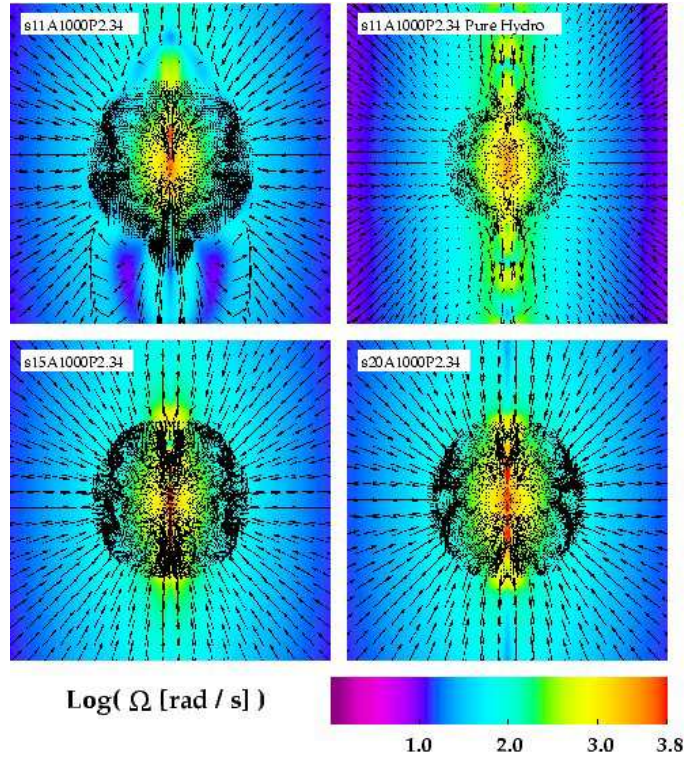} 
\caption{Logarithmic color map of the angular velocity
distribution in models s11A1000P2.34:2D, s11A1000P2.34:2Dh,
s15A1000P2.34:2D, and s20A1000P2.34:2D. Shown are the inner 400 km
on a side at 200 ms after bounce (150 ms for s11A1000P2.34:2Dh). 
Velocity vectors are superposed in the same way as
in Fig.~\ref{fig:s11_2Dpanel}. Model s11A1000P2.34:2Dh 
is evolved purely hydrodynamically. 
It explodes and leaves behind a less compact configuration
with smaller central $\Omega$ than the corresponding model
with MGFLD. With increasing progenitor iron core mass
(Table \ref{table:initialmodels}), the magnitude of the
final angular velocity increases and its distribution
becomes broader.
\label{fig:intermodel_2Dpanel}}
\end{center}
\end{figure}

In Fig.~\ref{fig:intermodel_2Dpanel},
we contrast the two-dimensional 
angular velocity distributions of models with different progenitor 
iron core mass and structure at $\sim$200~ms after core bounce. We
zoom in on the inner 400~km of our computational domain.
The rotation law parameters P$_0$ and A are prescribed. 
The top right panel of Fig.~\ref{fig:intermodel_2Dpanel} 
shows an $\Omega$ snapshot of model s11A1000P2.34:2Dh, 
the purely hydrodynamic variant
of s11A1000P2.34:2D. At $\sim$200 ms after bounce, 
this model has already undergone a
prompt explosion. This explains the outward pointing fluid 
velocity vectors in this panel. The region of high angular velocity is
stretched out along the axis of rotation (the vertical).
This is caused by the prompt explosion, which leads to
an ejection along the axis of high-$\Omega$ material from 
deep within the inner core.

\begin{figure}
\centerline{
\includegraphics[width=7.5cm]{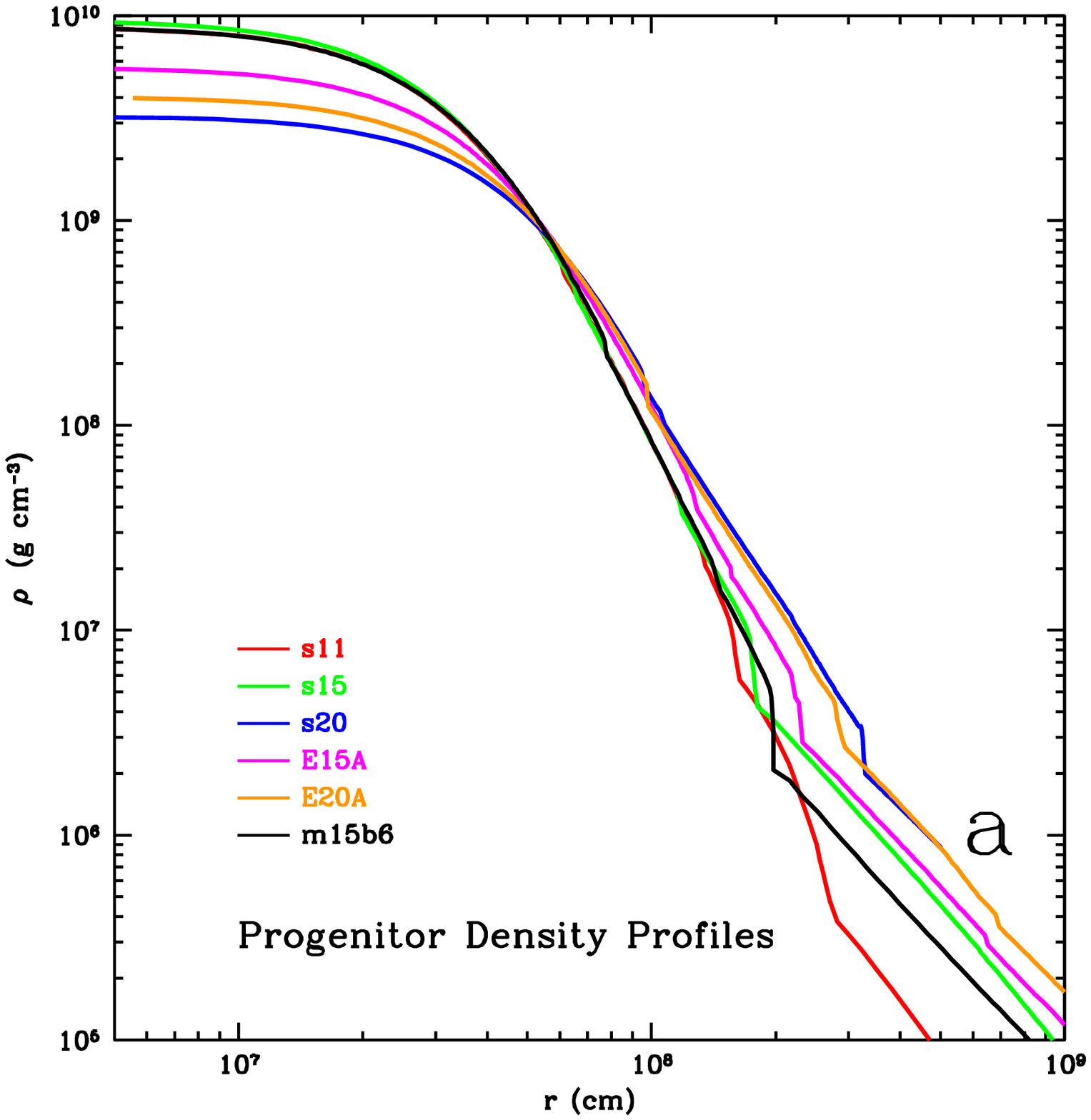}}
\caption{Initial mass density profiles for all progenitor models used in 
this study. s11, s15, s20 (11~\mo, 15~\mo, and 20~\mo, respectively)
are taken from \cite{ww:95}. Models E15A and E20A (15 \mo and 20 \mo)
are taken from \cite{heger:00} and model m15b6 is  
from \cite{heger:04}. E15A, E20A, and m15b6 were evolved in 
spherical symmetry with an approximate treatment of
centrifugal effects until the end of core carbon burning.
Note the great similarity in the density stratifications of models
s11, s15, and m15b6, especially interior to $\sim$2000 km. 
\label{fig:initialrhomass}
}
\end{figure}

The lower two panels of Fig.~\ref{fig:intermodel_2Dpanel}
show angular velocity distributions of models
s15A1000P2.34:2D and s20A1000P2.34:2D, respectively. Comparing them to 
the top-left panel which shows model s11A1000P2.34:2D, 
one notes that the PNS rotational structure is 
not strongly dependent on progenitor mass; the three models
in question have qualitatively and quantitatively very
similar angular velocity distributions at 200~ms after bounce.
For the s11 and s15 models, this is not surprising, given
that their iron core masses and density profiles
(Fig.~\ref{fig:initialrhomass} and Table~\ref{table:initialmodels}) 
are very similar to each other. However, the s20 progenitor model
which we use in our model s20A1000P2.34:2D has an appreciably different
initial iron core structure. It is more extended, has a lower
initial central density, and is significantly more massive than
the s11 and s15 models. Yet, at 200~ms after bounce only small 
differences in the angular velocity distribution are 
apparent. s20A1000P2.34:2D's PNS is more massive
and more extended. This is reflected in the slightly larger central
spheroidal bulge of high-angular-velocity material. As can be 
inferred from the flow patterns in Fig.~\ref{fig:intermodel_2Dpanel}, 
the thermodynamic structure and the resulting fluid dynamics 
of models s11A1000P2.34:2D, s15A1000P2.34:2D, and s20A1000P.23:2D 
are significantly different. These differences are  
reflected in the angular velocity distribution to only
a slight degree.

\begin{figure}
\begin{center}
\includegraphics[width=9cm]{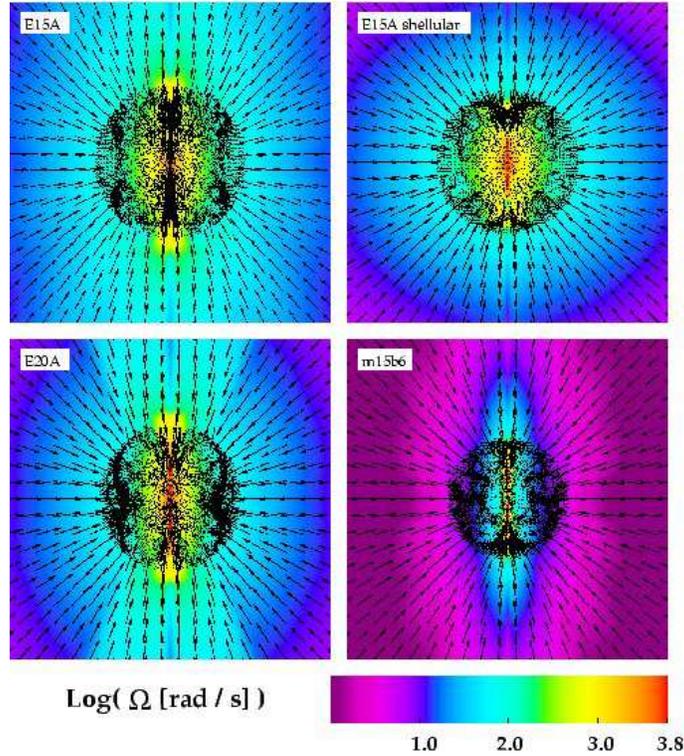} 
\caption{Logarithmic color map of the angular velocity
of models E15A:2D, E15A:2Ds (``shellular rotation"),
E20A:2D and m15b6:2D which use the rotating progenitors
of \cite{heger:00} and \cite{heger:04}. Shown are the
inner 400~km on a side at 200~ms after core bounce. 
Velocity vectors are superposed and downsampled at 
radii greater than about 200~km. 
The :2Ds variant of E15A is initially set up to
rotate with constant $\Omega$ on spherical shells. Its
region of high angular velocity is more compact 
than in its :2D counterpart, which is initially set up to rotate
with constant angular velocity on cylindrical shells. 
Interestingly, the innermost regions of E15A:2D and E15A:2Ds
are quite similar. Model E20A:2D has a more massive core than
model E15A:2D, but slower initial rotation. Its 
central angular momentum distribution is very similar to the
one seen in E15A:2D. Model m15b6:2D has a high initial period
of $\sim$32~s. 
\label{fig:hegermodel_2Dpanel}}
\end{center}
\end{figure}

Figure~\ref{fig:hegermodel_2Dpanel} depicts the angular velocity
distributions of the recent rotating ``Heger models'' 
(\citealt{heger:00}; \citealt{heger:04}) at 200 ms after core
bounce. We include models E15A, E20A, and m15b6 in our
two-dimensional model set and for model E15A:2D also perform
a simulation with initial ``shellular rotation'' (E15A:2Ds; 
see \S\ref{section:initialmodels}). The upper left and upper right
panels of Fig.~\ref{fig:hegermodel_2Dpanel} are snapshots of
the two-dimensional $\Omega$ distributions  
of models E15A:2D and E15A:2Ds, respectively. One can easily discern 
the more sphere-like $\Omega$ distribution of E15A:2Ds at large
radii. E15A:2Ds' angular velocity in the central region
exhibits a somewhat more spherical shape than that of its
E15A:2D counterpart. However, the magnitude of the angular velocities
observed does not depend strongly on the initial 
choice of rotation law, since the equatorial regions of the 
precollapse core that undergo the greatest spin-up during collapse have 
almost identical $\Omega$ before collapse (eq.~\ref{eq:rotlaw} and
the discussion in \S\ref{section:initialmodels}). The
bottom-left panel of Fig.~\ref{fig:hegermodel_2Dpanel} 
shows model E20A:2D's angular velocity snapshot at 200~ms after
bounce. Model E20A has an initial core mass and density profile 
that are similar to those of the s20 progenitor 
(Fig.~\ref{fig:initialrhomass}). 
Looking at Fig.~\ref{fig:initialrot},
one notes that E20A's initial $\Omega$ profile exhibits an
order-of-magnitude jump at around 1000 km, whereas E15A's
angular velocity jumps at about 3 times that radius. This
different behavior in the initial $\Omega$
profile and the initially longer central period
of the E20A progenitor (Table~\ref{table:initialmodels})
would naturally have led one to anticipate 
qualitative and quantitative differences in E20A:2D's angular
velocity evolution when compared with that of models 
E15A:2D and s20A1000P2.34:2D. In fact, E20A:2D exhibits an appreciably
different angular velocity profile. At lateral angles between
about 25 and 155 degrees, the angular velocity is almost
an order of magnitude lower than in the polar wedges. This
can be understood as follows:  At 200~ms after
bounce, the high-angular velocity material that was inside
a cylindrical radius of $\sim$1000~km has in the equatorial regions 
already accreted onto the PNS. What is then 
seen in the equatorial wedge is the angular velocity connected
with material of initial cylindrical radii above $\sim$1000~km
and, hence, of initially low angular velocity. However, in the polar region
we see material of large initial spherical radius,
but relatively small (below $\sim$1000~km) initial 
distance from the rotation axis. Hence, the higher angular velocity
seen at 200 ms in the polar wedge.

In the bottom-right panel of Fig.~\ref{fig:hegermodel_2Dpanel}
we display m15b6:2D's angular velocity distribution at 200~ms
after bounce. This model's initial central
iron core period is quite long (31.73~s) and its iron core
structure is similar to that of the older s11 and s15 progenitors 
(\citealt{ww:95}). Its iron core angular velocity profile
exhibits the jumps at compositional boundaries 
which are characteristic of the set of rotating
progenitors considered here. Again, arguing that the initial
rotational profile is determining the PNS spin,
we compare m15b6:2D with s11A1000P41.89:2D (bottom-right
panel in Fig.~\ref{fig:s11_2Dpanel}). Even though model
m15b6:2D has a shorter initial central period than 
s11A1000P41.89:2D, its central spheroidal 
bulge of high-angular velocity is smaller
at 200~ms after bounce. A glance at Fig.~\ref{fig:initialrot}
reveals the cause: m15b6's iron core is indeed spinning
faster than s11A1000P41.89, though only in the inner
$\sim$700~km of the precollapse iron core. At that radius,
m15b6's $\Omega$ drops below the angular velocity in model 
s11A1000P41.89:2D due to the use of the rotation law of 
eq.~(\ref{eq:rotlaw}) in the latter model. 

\begin{figure}
\begin{center}
\includegraphics[width=7.5cm]{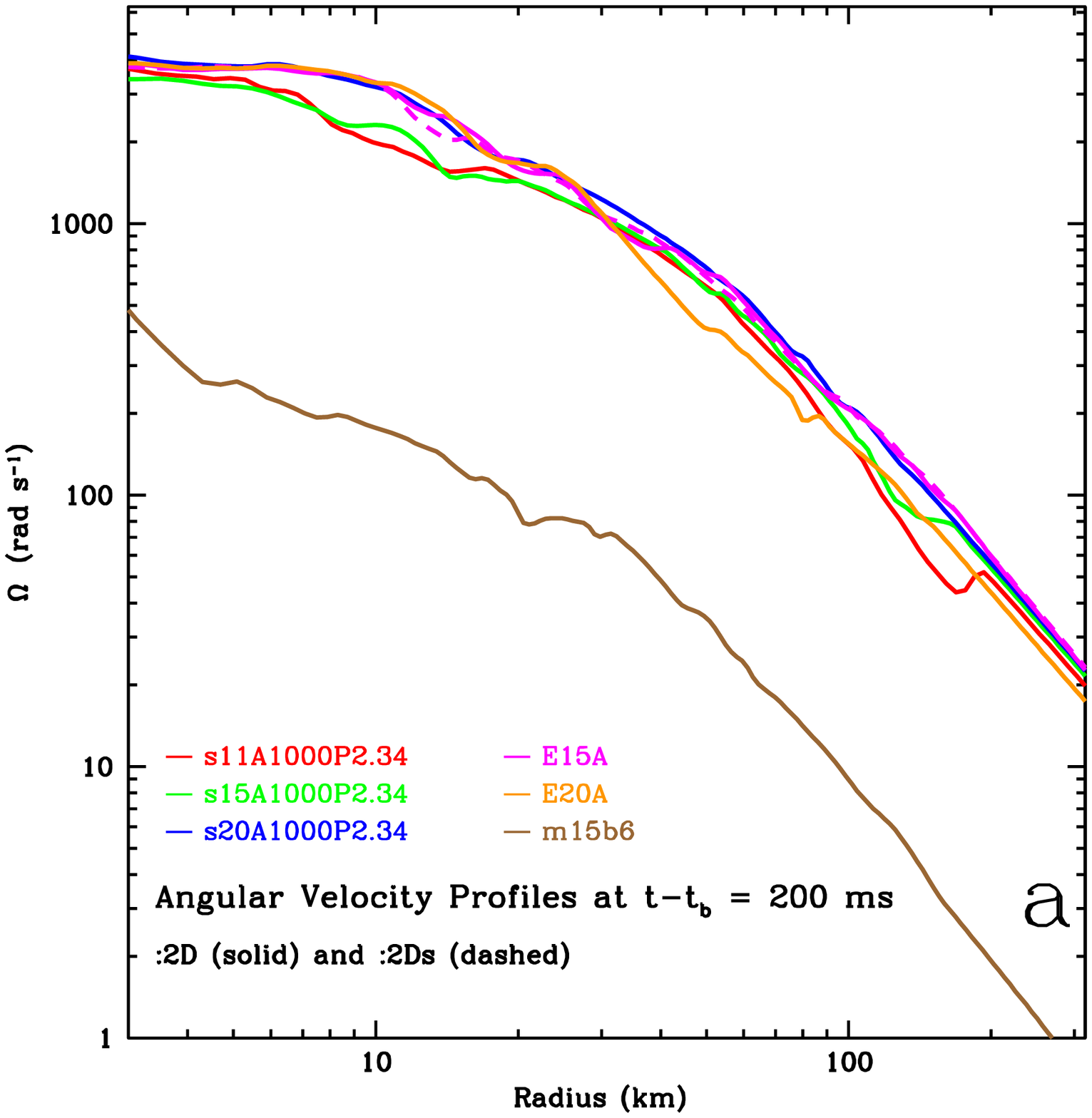}
\includegraphics[width=7.5cm]{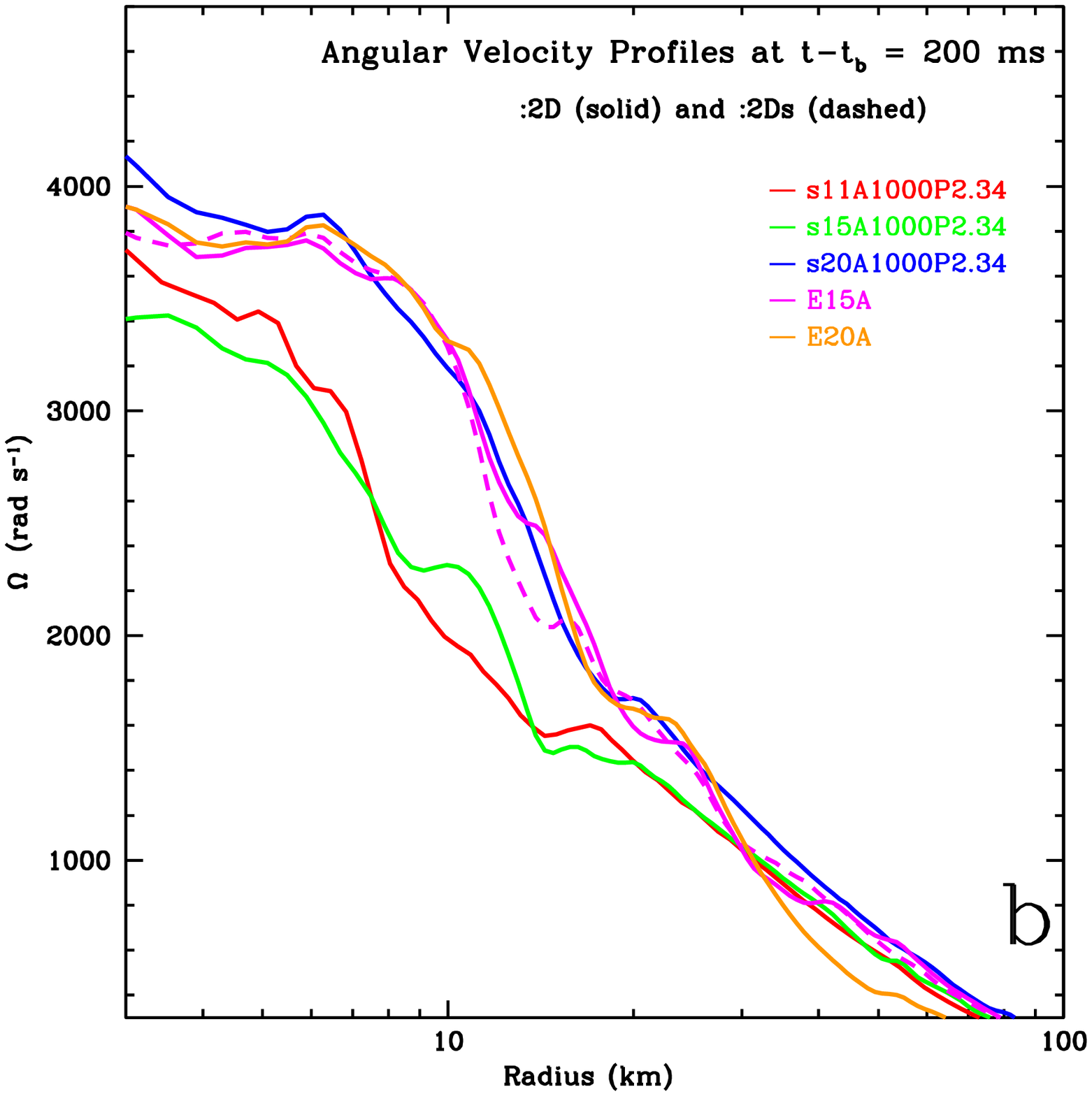}
\caption{Equatorial profiles of the angular velocity at 200 ms after
bounce for a set of models with similar initial iron core periods
(P$_0$) and different progenitor mass. The slow ``magnetic" 
model m15b6 (\citealt{heger:04}) stands out and has been included in
(a) for comparison. (a) shows $\Omega$ on a logarithmic
scale vs. radius and (b) displays $\Omega$ on a linear scale
for the innermost 100 km of each model. On a logarithmic scale
it is hardly possible to distinguish between individual fast models. 
With the exception of s11 and s15 that have very similar progenitor 
structure, more massive progenitors lead to slightly 
higher PNS central $\Omega$. The model E15A:2Ds 
(dashed-magenta graph) is setup in ``shellular" rotation. 
This, however, does not lead to 
a considerably different equatorial $\Omega$ profile 
at postbounce times than that of model E15A:2D. Differences 
in off-equatorial regions are, however, present (see 
Fig.~\ref{fig:hegermodel_2Dpanel}).
\label{fig:intermodel_2D_200_omega}}
\end{center}
\end{figure}

Figure~\ref{fig:intermodel_2D_200_omega} shows equatorial
angular velocity profiles at 200 ms after bounce.
The figure contains a panel with $\Omega$ on a logarithmic
scale out to 300~km in equatorial radius and a more zoomed-in
version showing only the inner 100~km. $\Omega$ is
drawn on a linear scale to more clearly display the differences
between the individual models. We contrast the equatorial profiles of 
models s11A1000:2D, s15A1000:2D, and s20A1000:2D 
with P$_0$ set to 2.34~s with the Heger models
E15A:2D, E20A:2D, and m15b6:2D. Above, we discussed the systematic
changes of the PNS spin profile due to 
variations in the initial rotational configuration for a
fixed progenitor structure. In the discussion of the
two-dimensional angular velocity snapshots, we saw
that progenitor mass and iron core structure 
appear to have only a secondary effect on the PNS's 
birth spin rate and rotational
configuration. This is supported by the equatorial rotational
profiles in Fig.~\ref{fig:intermodel_2D_200_omega}. 
Models with similar initial (quantitative) 
rotational profile yield similar PNS rotational
profiles, despite significant differences in their 
progenitor structures. Only in the linear-scale panels does one
note differences that can be attributed to differences
in the initial angular velocity distribution (``spherical'' vs.
``cylindrical'' and rotation law vs. one-dimensional rotation). 
For example, we noted the interesting shape of E20A:2D's angular
velocity distribution in Fig.~\ref{fig:hegermodel_2Dpanel}.
The early drop of $\Omega$ in the equatorial regions is reflected
in the dropoff in equatorial angular velocity
at about 30~km (clearly visible in the right panel
of Fig.~\ref{fig:intermodel_2D_200_omega}). Note that the initial
discrete jumps of the angular velocity of models E15A:2D,
E20A:2D, and m15b6:2D are naturally smoothed out during the 
two-dimensional collapse and postbounce evolution. 

\begin{figure}
\begin{center}
\includegraphics[width=7.5cm]{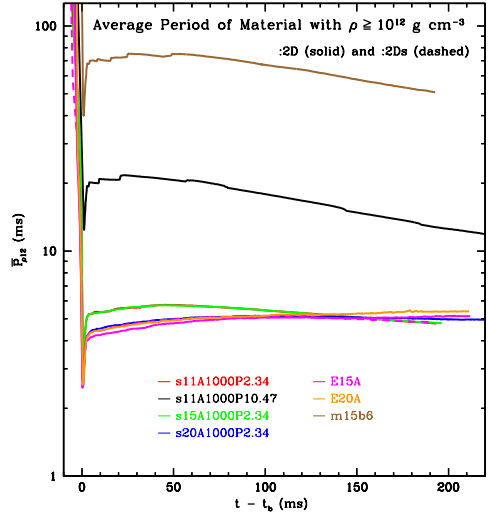}
\caption{Postbounce evolution of the moment of inertia-weighted
mean period \pav for models with differing initial
iron core spin and progenitor mass. \pav strongly depends on the
initial iron core spin and weakly on progenitor mass and structure.
Interestingly, the choice of rotation law (``cylindrical" vs. 
``shellular" --- see dashed plot for E15A:2Ds) has little effect on the 
mean period.  The small discontinuities in some of the graphs
at early postbounce times are due to the finite resolution of 
our numerical grid and occur when an entire mass shell first exceeds 
the density threshold ($\rho$~$\ge$~10$^{12}$~g~cm$^{-3}$). 
\label{fig:intermodel_2D_periods}}
\end{center}
\end{figure}

In Fig.~\ref{fig:intermodel_2D_periods}, we compare the 
moment-of-inertia-weighted mean period \pav of models with different 
progenitor structures. For comparison, we enlarge the sample 
of models to include s11A1000P10.47:2D and E15A:2Ds.
The evolution of \pav supports our assertion that progenitor
structure has only a minor effect on the spin period
of neutron stars at birth in our two-dimensional models. 
The magnitude of initial angular velocity in the part of the 
precollapse iron core that forms the homologous inner core 
during collapse is the 
most important parameter for the PNS spin.
However, this assertion holds only for the early stages
of the PNS evolution. It is possible that
the rotational configuration could be influenced by, 
among other phenomena, late-time 
fall-back accretion whose rate and absolute
accreted mass will depend on the progenitor
mass density profile. Furthermore, black hole formation
will become more probable with increasing progenitor mass,
leading to a totally different scenario, possibly to a 
collapsar-type GRB (\citealt{woosley:93,macfadyen:01}).

\begin{figure}
\begin{center}
\includegraphics[width=7.5cm]{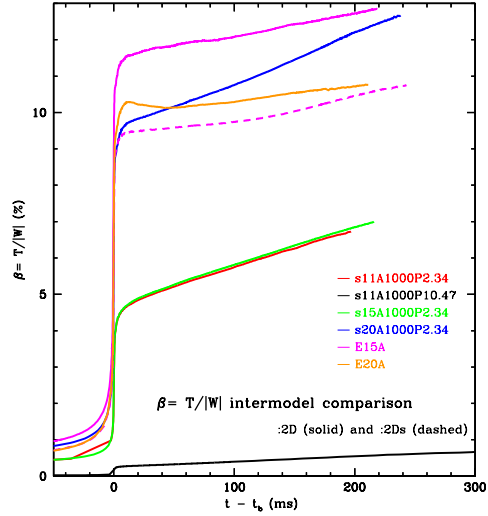} 
\caption{Evolution of the rotation parameter $\beta$~=~T/$|$W$|$.
Here, we compare a representative set of models from our 
two-dimensional model
series. Due to the similarity of their progenitor structures,
s11A1000:2D and s15A1000:2D have almost identical $\beta$ evolutions
for the same initial iron core spin period (red and green graphs). 
$\beta$ increases linearly after bounce in all \cite{ww:95} models.
Models E15A:2D, E15A:2Ds, and E20A:2D have considerably more 
complicated rotational structures (Fig.~\ref{fig:initialrot}). 
Hence, an appreciably different $\beta$ evolution can be expected 
(T $\propto$ $\Omega^2$). Model s20A1000P2.34 shows the steepest
$\beta$ slope and is likely to reach the classical limit for secular
rotational instability ($\beta$ $\ge$ 14 \%). However, a low-T/$|$W$|$
instability might set in at much lower $\beta$ (\citealt{rotinst:05}).
\label{fig:intermodel_2D_beta}}
\end{center}
\end{figure}

To conclude this section on our 2D axisymmetric
models, we discuss the evolution of the rotation parameter
$\beta$ for different progenitor masses. In
Fig.~\ref{fig:intermodel_2D_beta}, we compare the 
postbounce $\beta$ evolution for a representative set of
models. $\beta$ is an integral quantity, evaluated by
integrals that encompass all grid points. Differences 
(and similarities) in the $\beta$ evolution caused by progenitor 
structure and choice of initial angular velocity distribution and
mapping are obvious. Models s11A1000P2.34:2D and 
s15A1000P2.34:2D are evolved with the same initial rotation
law and start with identical P$_0$. Their iron core structure
is very similar as well. Their $\beta$ behavior turns out to be almost
identical. Model s20A1000P2.34:2D is evolved using the
same initial rotational configuration, but has a significantly
more massive iron core with an initially lower central density.
Its higher value of $\beta$ at bounce and during its postbounce
evolution is a consequence of the much greater angular momentum
contained in s20A1000P2.34:2D's core and of its relatively high 
initial value of $\beta$.
In Table~\ref{table:results2D}, we list the total angular
momentum contained in material above 10$^{12}$~g~cm$^{-3}$ at the
end of each model's numerical evolution. 
The high value of $\beta$ at bounce is also in part a 
consequence of its initially lower 
core density. This results in a greater spin up of the inner
core material during collapse. Model E20A:2D's initial central
spin period is a bit shorter than s20A1000P2.34:2D's (2~s vs. 2.34~s),
while its initial inner core density is slightly larger.  
E20A:2D's $\beta$ is not much larger at bounce
than s20A1000P2.34:2D's, but starts increasing
with a greater rate postbounce. However, that increase
is stopped about 10~ms after bounce when the initial profile's 
big jump in $\Omega$ at about 1000~km 
(Fig.~\ref{fig:initialrot}) reaches small radii and the increase
of rotational kinetic energy is abruptly slowed. Qualitatively,
and to some extent quantitatively, this behavior can be inferred 
from the two-dimensional angular velocity snapshot  
presented in Fig.~\ref{fig:hegermodel_2Dpanel}.

Overall, E15A:2D is the fastest
model of our two-dimensional model set. Its initial central
period is as low as 1.5~s and the {\it initial} value of $\beta$
amounts to 0.5\%. Its initial inner core density is considerably
lower than for the slower or nonrotating 15 \mo\ models 
(Fig.~\ref{fig:initialrhomass}). It does, like E20A:2D, exhibit a 
discontinuity in $\Omega$ in its initial profile. However, 
this discontinuity is located at greater radii and does not advect 
in before the end of our simulation. All this explains the 
high $\beta$ at bounce and the postbounce increase seen 
for this model.  However, understanding what leads to the 
differences between
models E15A:2D and E15A:2Ds requires a bit more involved discussion.
In ``shellular'' rotation, the iron core angular velocity is set up
to decrease roughly with $r^{-2}$, where $r$ is the spherical radius 
(see eq.~\ref{eq:rotlaw} and \S\ref{section:initialmodels}). 
In comparison with ``cylindrical''
rotation, off-equatorial grid points at a given distance $\varpi$ from
the rotation axis are set up with an initial angular velocity
that is reduced by a factor of $\varpi / \sqrt{\varpi^2 + z^2}$,
where $z$ is the cylindrical coordinate on the symmetry axis. Hence,
there is less total angular momentum on E15A:2Ds' grid and its 
initial value of $\beta$ is about 30\% smaller than E15A:2D's. 
These facts determine this model's $\beta$ evolution.

As expected, model E15A:2D reaches the greatest postbounce 
$\beta$ of the entire set of two-dimensional models presented here. 
At the end of the calculation, $\beta$ is 12.85\% and is still 
increasing. During their late-postbounce evolution, models 
E15A:2D and s20A1000P2.34:2D and, perhaps, models E20A:2D and E15A:2Ds
are likely to reach the classical threshold for secular 
rotational instability. However, it is unlikely that any of the
models presented here will ever reach the threshold $\beta$ at which 
classical high-T/$|$W$|$ rotational instabilities
would set in. Nevertheless, there are multiple models that may 
become unstable to the low-T/$|$W$|$ instability of postbounce cores 
observed by
\cite{rotinst:05}.

\section{Comparison of One- and Two-Dimensional Simulations}
\label{section:comp1D2D}

\begin{figure}
\begin{center}
\includegraphics[width=14cm]{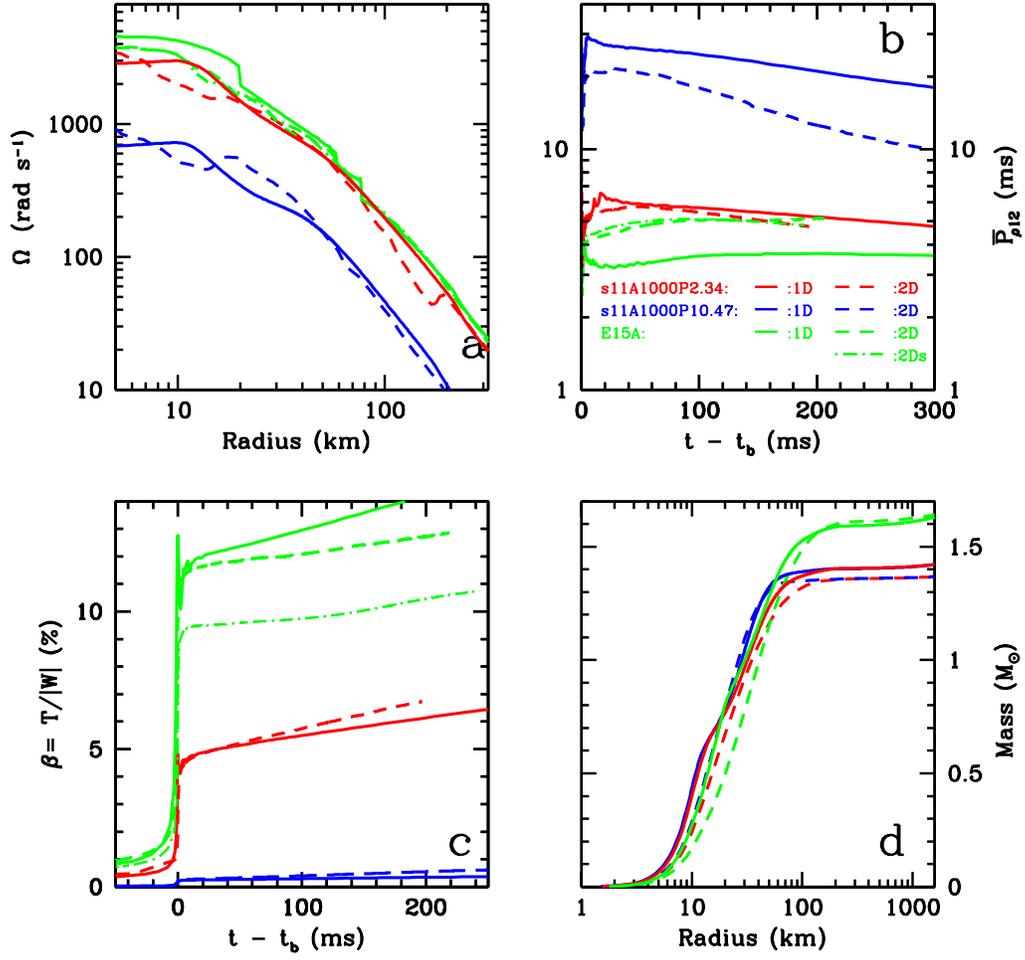} 
\caption{Contrasting one-dimensional to two-dimensional results:
In panel~(a), the equatorial angular velocity profiles at 
200~ms after bounce of three models, namely s11A1000P2.34~(red), 
s11A1000P10.47~(blue), and E15A~(green), each in spherical 
symmetry~(solid lines) and axisymmetry~(dashed lines), are compared. 
The profiles are plotted from 5~km on outwards to excise axis
artefacts. For E15A:2D, we also plot the ``shellular" variant.
Panel~(b) shows the evolution of the moment-of-inertia weighted
mean period \pav for the selected set of models. In 
Panel~(c), we display the evolution of the rotation parameter
$\beta$ and panel~(d) depicts the mass-(spherical)radius relationship
 at 200 ms after bounce. 
\label{fig:comp1D2D_panel}}
\end{center}
\end{figure}

In \S\ref{section:results1D} and \S\ref{section:results2D}, we 
presented the results
of our one- and two-dimensional simulations separately. 
Here, we compare the two sets of simulations. 
Panel (a) of Fig.~\ref{fig:comp1D2D_panel} shows the 
{\it equatorial} angular velocity profiles at 200~ms after bounce
for models s11A1000P2.34, s11A1000P10.47, and E15A. Two-dimensional
variant (:2D) profiles are the dashed lines, while
one-dimensional variants (:1D) are the solid lines. For model
E15A we also plot the version with ``shellular'' rotation (:2Ds).
First, we note that the rotational profiles of 
E15A:2D and E15A:2Ds lie practically on top of each other. This
is remarkable, but not entirely surprising if one recalls 
that the initial equatorial angular velocity profiles of these
models are identical. However, Fig.~\ref{fig:hegermodel_2Dpanel}  
demonstrates that there are significant differences once one leaves
the equator. 

Comparing the :1D profiles to their :2D counterparts,
we note that there is good agreement outside $\sim$50~km. 
With decreasing radius the :1D and :2D profiles for a given model 
begin to diverge. The two-dimensional models have less smooth profiles.
The substructure in the equatorial profiles is created by 
angular momentum advection in convective plumes
inside the protoneutron star and in the postshock region. Since
convection is confined to small cylindrical radii in quickly
spinning PNSs, the equatorial profiles of faster models overall are  
smoother and closer to those of their :1D counterparts 
than those of slow models that allow for more vigorous convection 
at greater cylindrical radii. Note that the jump in $\Omega$ visible 
in the :1D profile of E15A (green solid-line, at $\sim$20~km, also see 
Fig.~\ref{fig:initialrot}) is smoothed out in its 
:2D counterpart. 

In panel~(b) of Fig.~\ref{fig:comp1D2D_panel}, we display
the moment-of-inertia-weighted mean period \pav for the same
set of models. \pav is an integral quantity based on a
density criterion which takes into account the greater 
dimensionality of the :2D models. Model E15A is the fastest
model for which we have :1D and :2D simulation data. Its initial
central period is 1.5~s and it undergoes core bounce under 
strong centrifugal influence. The postbounce \pav evolutions 
of E15A:2D and E15A:2Ds differ significantly from those of their
one-dimensional counterparts. At and after bounce, the two-dimensional
PNSs are able to relax to greater PNS radii and energetically more
favorable configurations than the spherically symmetric model. Hence,
the :2D \pav of this (almost) rotationally supported model is 
larger than the :1D's. With increasing initial rotation period, 
rotational support for the PNS becomes less and less and the
overall greater angular momentum in the :2D models leads to shorter
PNS \pav$\mathrm{\!s}$ (see \S\ref{section:initialmodels}; ``cylindrical'' vs.
``shellular'' rotation; also see the discussion towards the
end of \S\ref{section:results2D}).

The evolution of the rotation parameter $\beta$ is shown
in panel (c) of Fig.~\ref{fig:comp1D2D_panel}. Again, there
are significant differences between :1D and :2D variants.
For E15A, $\beta$ is overestimated by the one-dimensional model and 
for the slower models s11A1000P2.34 and s11A1000P10.47 the 
one-dimensional models underestimate $\beta$. Model E15A:2Ds which
is in ``shellular'' rotation reaches lower $\beta$ than E15A:2D 
because there is less total angular momentum in the former. The
explanation for the differences in $\beta$ between one-dimensional
and two-dimensional models is analogous to the explanation for the 
differences seen in the \pav evolution discussed above.

In panel (d) of Fig.~\ref{fig:comp1D2D_panel}, we display
the mass-(spherical)radius relationship for selected :1D and :2D models. 
Here, the naive first assertion holds: the slower a :2D model is, 
the more similar is its mass distribution to that of the corresponding 
:1D model and the better the :1D and :2D mass profiles agree. However, 
keep in mind that for fast :2D models that experience significant 
rotational flattening a mapping $M(r)$ will always be ambiguous.

We conclude that the overall characteristics and systematics
of the mapping between initial model and rotational profile
to the postbounce spin of the PNS are rendered with acceptable 
accuracy by the one-dimensional treatment of rotation. 
There are, however, significant qualitative and quantitative
differences between :2D and :1D in all models.  In particular,
we {\it do not} expect convergence of the :1D and :2D results
with increasing P$_0$ since multi-dimensional effects, especially
convection that becomes more vigorous in slower models, will
always distinguish :2D evolution from :1D.

\section{Simple Estimates \\ of the Final Neutron Star Spin}
\label{section:simpleP}

The simulations presented in this paper cover only the very early
stages of a neutron star's life. During tens to hundreds 
of seconds after bounce and a successful explosion, 
the neutron star will deleptonize and cool via neutrino emission. 
For the typical neutron star mass range of $\sim$1.2-1.6 \mo\ 
(\citealt{fryerkalogera:01}; \citealt{thorsett:99}), 
\cite{lattimerprakash:01} find a radius of
$\sim$12~km and a fiducial moment of inertia
of $I$ = 0.35 $M R^2$ for a cold neutron star. 
Cold neutron stars are likely to be in solid-body 
rotation since this is the energetically most favorable configuration
and viscous processes will most likely drive the neutron star into rigid
rotation during its cooling phase. 

A number of studies of stellar evolution 
(\citealt{heger:00, hegerwoosleyspruitlanger:03, heger:04})
and of rotating core-collapse and supernova evolution 
(\citealt{fryerheger:00,fryerwarren:04}) have estimated the final
neutron star spin period via  
\begin{equation}
\label{eq:simpleP}
P_{NS} = 2\pi \frac{1}{(J / I)}\,\, .
\end{equation}
Here $J$ is the supposed final total angular momentum and $I$
is the moment of inertia obtained with the above prescription
and a choice of neutron star mass and radius (typically
$M = 1.4$~\mo\ and $R$=12~km). For comparison with other
studies, we estimate the asymptotic final neutron 
star spin of our models in the following
manner.  We consider our two-dimensional simulations 
and a representative subset of our one-dimensional simulations
and choose $M$ to be the mass of all material with a
density above 10$^{12}$~g~cm$^{-3}$ at 200~ms after bounce.
We choose this time to ensure that we have one- and two-dimensional
data for our estimates. Also, that this is a sensible choice becomes 
apparent when looking at Fig.~\ref{fig:intermodel_2D_mass}, 
which shows the mass interior to a given spherical radius at 
about 200~ms after bounce for representative models. 
$M_{\rho 12,200}$ encompasses the majority of the total mass on the 
grid, located interior to some tens of kilometers. We assume that a 
successful explosion removes the remaining matter and that 
$M_{\rho 12,200}$ is the final neutron star mass. For simplicity, 
no further fall-back accretion or neutrino losses are considered. 
For the angular momentum in eq. (\ref{eq:simpleP}), we choose 
$J_{\rho 12,200}$, which is the total angular momentum of mass 
$M_{\rho 12,200}$. 
We assume this angular
momentum and mass to be conserved during the contraction 
to a final radius of 12~km. 
In addition, we compute the neutron star's
rotation parameter $\beta_{NS}$ = T/$|$W$|$ 
and the ratio of the neutron star's angular velocity to the
Keplerian break-up velocity, defined by 
$\Omega_K = \sqrt{GM/r^3}$. For $\beta$, we
estimate the gravitational potential energy according to the
prescription given in \cite{lattimerprakash:01}:
\begin{equation}
\label{eq:w}
|\mathrm{W}| \simeq Mc^2 \times 0.6 \frac{GM}{Rc^2}  / 
\big(1 - 0.5 \frac{GM}{Rc^2}\big)\,\, .
\end{equation}
We summarize our estimates in Table~\ref{table:simpleP}.
Neutron stars with $\beta$ > 50\% (Viral theorem limit) 
and angular velocity above the Keplerian break-up velocity cannot 
exist. For six of the two-dimensional models listed
in Table~\ref{table:simpleP} our estimate leads to a spin rate
beyond break-up and to unphysically high values of $\beta$. 
Since the angular momentum contained in one-dimensional models
is smaller than in the :2D models, only the two most extreme :1D
models reach such high $\beta$ and $\Omega/\Omega_K$.
These models simply cannot form stable compact 
neutron stars, unless there are one or multiple mechanisms that 
remove angular momentum from the cooling neutron star. Otherwise,
the neutron star will cool, but remain centrifugally hung-up 
with a disk of matter at Keplerian velocity.
Two-dimensional models with initial central iron core
spin period above $\sim$5~s may form stable neutron stars
below the Keplerian limit (s11A1000P4.69:2D, s11A1000P10.47:2D,
s11A$\infty$P10.47:2D,s11A1000P41.89:2D,s11A1000P157.10:2D, and
m15b6:2D). For the one-dimensional models, this initial period
cut is somewhat lower, allowing iron cores with initial periods
of down to $\sim$2~s to form stable neutron stars. However, keep
in mind that the one-dimensional description of rotation is only
approximate and underestimates the angular momentum in the PNS 
(see \S\ref{section:comp1D2D}).

Only the overall slowest two-dimensional model
(s11A1000P157.10:2D) produces a neutron star with a period 
in the ballpark of observed young pulsar periods 
(tens of milliseconds). Models s11A1000P41.89:2D and m15b6:2D 
yield neutron stars with $\sim$7~ms period. This number
is in rough accord with the periods derived by 
\cite{hegerwoosleyspruitlanger:03} for progenitors with
similar initial rotation period and angular momentum distribution. 
However, \cite{heger:04} estimated a neutron star period of 11~ms for
their m15b6 progenitor (cf.~m15b6:1D's P$_{NS}$ $\sim$~14~ms). 
\cite{fryerwarren:04} estimated for E15A and for a m15b6-like model 
neutron star periods of 0.91~ms and 17~ms,
respectively. However, we note that in their three-dimensional
simulations the PNS appears to continuously lose
angular momentum after bounce. These authors 
do not provide a conclusive explanation for this phenomenon.

We have surveyed the parameter space of initial central
iron core periods from $\sim$1.5~s to $\sim$160~s with one- and
two-dimensional supernova simulations.
Across the board, our estimates of the final neutron 
star spin periods, based on the two-dimensional simulations
that handle rotation in a consistent manner, are shorter than found in  
previous studies. A large fraction of the two-dimensional models
considered in this study produce cold, contracted neutron stars
beyond the break-up spin rate. Given that observed young pulsar periods 
are between tens and hundreds of milliseconds, 
we conclude that there must be robust processes that  
efficiently spin down a neutron star before it emerges 
as a pulsar {\it or} that progenitor iron cores generically rotate 
more slowly than many of our progenitor models.  
We discuss in the following subsections a number of physical 
processes which could carry away or redistribute a significant 
amount of the angular momentum of the nascent neutron star.

\subsection{Viscous Processes}

In this paper, we have already considered a mechanism for 
spindown of PNSs: 
viscous angular momentum redistribution 
(by, e.g., the MRI)\footnote{Other processes, such as the viscosity 
associated with convective turbulence, may also operate on the shear 
energy stored in differential rotation (Thompson et al.~2005)}. 
Because rotational energy scales with $\Omega^2$, viscous heating
scales with $\Omega^3$, and the viscous timescale is proportional
to $\Omega^{-1}$, the multiple effects of viscosity are most 
noticeable in fast rotators (\citealt{thompvisc:05}). 
However, even for our rapidly rotating models (P$_0$~\sless~4~s), 
our results (see \S\ref{section:results1D} and 
Tables~\ref{table:results1D} and~\ref{table:simpleP}) 
indicate that viscous effects are not able to increase the spin 
period by more than a factor of 2~--~3 
on 100~ms to 1~s timescales. Although potentially interesting, these
processes are not sufficient to yield fully-contracted 
10--20~ms neutron stars.   For slower initial iron core spin periods 
(P$_0$~\sgreat~4~s), viscous effects quickly become much 
less important on the timescales considered here.

\subsection{Secular and Dynamical Rotational Instabilities}

PNSs that spin with high enough $\beta$ might undergo
the low-T/$|$W$|$ instability (for T/$|$W$|$ \sgreat~8\%,
\citealt{rotinst:05}) or the classical secular and dynamical
rotational instabilities at higher T/$|$W$|$. These instabilities will
put a lower limit on the neutron star spin of $\sim$1 ms 
(set by the value of $\beta$ at which the star regains stability, which
in turn depends on the degree of differential rotation and the
equation of state), but cannot lead to a significant spindown 
of the neutron star.

\subsection{r-Modes/Gravitational Wave Emission}

Rossby-waves (r-modes) in rapidly rotating neutron stars 
are driven unstable by the emission of gravitational radiation.
Consequently, stellar rotational energy is converted into both
gravitational waves and r-mode energy 
(\citealt{andersson:98,lindblometal:01}). The spindown
torque presented in \cite{arras:03} is
\beq
{\cal M} \simeq (2\times10^{49} \mathrm{dyn\,cm})
\bigg(\frac{\alpha_e}{0.1}\bigg) \nu^{12}_{KHz}\,\,,
\eeq
where $\alpha_e$ is the saturation amplitude of the r-mode 
in dimensionless units and $\nu_{KHz}$ is the neutron star
spin frequency (1/P) in units of kHz. With this, we estimate
\beq
\dot{\mathrm P} = \frac{1}{2\pi} \frac{\cal M}{I} {\mathrm P}^2\,\,.
\eeq
Because of the low saturation level, $\alpha_e$~$\sim$0.1 
(\citealt{arras:03}), and the extreme spin-frequency dependence,
r-modes can lead to a factor of two increase within 
a matter of days in the period of neutron
stars that spin close to break-up, but
will require millennia to slow down moderately rotating neutron stars
(\citealt{arras:03}; \citealt{heger:04}).

\subsection{Neutrino Emission}

Neutrinos carry away $\sim$99\% of the gravitational binding energy
of a forming neutron star. Besides the possibility 
of a net linear momentum flux,
which could be responsible for the observed neutron star
birth kicks, neutrinos also carry away angular momentum
(\citealt{epstein:78,baumgarteshapiro:98,jankaproc:04}).
Assuming that the neutrinos are diffusing out of the neutrinosphere
in all directions, \cite{jankaproc:04} estimates that a maximum
of $\sim$40\% of the total angular momentum can be lost if
the axis ratio of the PNS is close to one.
Hence, this mechanism could --- in the most optimistic case ---
lead to a factor-of-two spindown for a relatively slow 
PNS (\pav~\sgreat~10~ms).

\subsection{Rotation-Powered Supernovae}

The typical supernova explosion kinetic energy is near 10$^{51}$~erg.
The rotational energy stored in the PNSs in
our models is easily calculated using T~$= 1/2\, J_{\rho,12}\, 
(2\pi/$\pav).
Model s20A1000P2.34:2D has a rotational energy
of $\sim$2~$\times$~10$^{52}$ erg. If only 10\% of this model's
rotational energy is converted into radial kinetic energy, this
could power the entire supernova explosion, as suggested in
the ``jet-driven'' supernova mechanism (\citealt{akiyama:03}; 
\citealt{akiyama:05}).
In that picture, the PNS represents the fly-wheel
that provides the rotational energy to drive (magneto-) hydrodynamic
jets along the poles leading to a bipolar supernova explosion.
If 10\% of the rotational energy is dumped into the supernova, the
PNS will spin down by not more than $\sim$5\%. 
If, however, 50\% of the rotational energy is imparted to the supernova, 
a spindown by $\sim$30\% results.

\subsection{Magnetic Protoneutron Star Winds}
\label{magneto}

In the seconds after collapse and explosion,
a thermal non-relativistic neutrino-driven wind emerges from the 
PNS (Duncan et al.~1986). 
If the PNS has a magnetar-strength 
(\sgreat~$10^{14}-10^{15}$ G; Duncan \& Thompson 1992) 
large-scale surface magnetic field,
then the wind material will be forced to corotate with the
PNS.  For rapid rotation, the wind is 
magneto-centrifugally whipped off of the stellar surface 
and the neutron star's angular momentum and rotational 
energy may be efficiently extracted 
(\citealt{thompmagnetar:04,thompapjl:03}).
The spin period increases as 
$\dot{\mathrm P}/{\mathrm P}\approx3(\dot{M}/M)(R_A/R_{NS})^2$,
where $\dot{M}$ is the mass loss rate and $R_{A}$ is the
Alfv\'en radius, where the kinetic energy density of the 
wind first exceeds the magnetic energy density. If the wind carries
away angular momentum at a constant rate over a time $t$, then
the relative change in period is 
P/P$_{\rm birth} \sim \exp(\mathrm{\dot{P}}t/{\mathrm P})$.
Assuming a 1.4~\mo\ PNS with a 2~ms period, 
a radius of 20~km, an Alfv\'en radius of 40~km, and a mass-loss
rate of 10$^{-3}$ \mo\ s$^{-1}$ (\citealt{thompmagnetar:04}), together 
with a 10~s wind duration, we obtain 
$\Delta {\mathrm P}$/$\mathrm{P}_{birth}$ of 
order 10\%. Although this mechanism does not spin the PNS down 
significantly on a 10 s timescale, the non-relativistic wind 
carries of order 10$^{51}$~erg in kinetic energy.

As any PNS cools and contracts, the neutrino luminosity
decreases and the mass loss rate abates. As a consequence, 
for fixed magnetic field strength, the wind becomes 
Poynting-flux dominated and relativistic. For near-millisecond 
spin periods and magnetar-strength fields, this transition happens 
just seconds after the explosion commences
(\citealt{thompmagnetar:04}).

Assuming a dipolar field configuration --- probably
pessimistic considering the fact that pulsar braking
indices are observed to be less than 3 (and, see
Bucciantini et al.~2005) --- the rotational energy loss 
rate is $\dot{E}\approx B^2 R^6 \Omega^4/c^3$ so that the spindown
timescale is just P$/\dot{\mathrm P}\sim1.7\times10^4$~s for a 
protoneutron star with $R$~=~20~km, $B$~=~10$^{14}$ G, P~=~2~ms, 
and moment of inertia $I$~=~0.35M$R^2$.  For these parameters, 
essentially all of the rotational energy can be lost to the 
relativistic wind in just tens of hours.

The two most obvious and fundamental problems with spinning
down a millisecond PNS rapidly with non-relativistic or relativistic 
pulsar-like winds is that (1) 
normal neutron stars are not expected to be born with
large-scale magnetar-strength fields and (2) if the PNS
is efficiently spun down, the remnant should bear the signature
of the $\sim$10$^{52}$~ergs of rotational energy deposited 
within it. However, observations of generic young SNRs do not show
evidence of such energy deposition.

\subsection{Fall-Back / Propeller Mechanism}

After a successful explosion there might be a phase during which the
neutron star accretes a considerable amount of material that has 
failed to reach escape velocity in the shock-driven expulsion of the
stellar envelope. \cite{macfadyen:01} estimate the late-time 
($\ge$1000~s after explosion; \citealt{heger:04}) 
fall-back accretion rate, based on the parameters of SN~1987A, to be 
\beq
\dot{M} \equiv 2 \times 10^{26}\,\, t_5^{-5/3} \mathrm{g}\,\,  
\mathrm{s}^{-1}\,\,,
\eeq
where $t_5$ is the time in units of 10$^5$~s.
If the then deleptonized, but possibly centrifugally-hung-up
neutron star has a sufficiently strong magnetic field, its Alfv\'en
radius can be greater then the radius at which the
corotation angular velocity exceeds the Keplerian orbital speed.
The infalling material is halted by magnetic forces at the
Alfv\'en radius, spun up to corotation and, if $\Omega$ > $\Omega_K$,
expelled. The ejected matter carries away angular momentum and 
leads to a spindown of the neutron star. \cite{heger:04}, based
on the work of \cite{alpar:01} (also see \citealt{romanova:04}
for recent numerical results), estimate that this so-called
propeller mechanism can very efficiently spin down neutron stars
with surface magnetic field strengths above $\sim$2$\times$10$^{13}$~G.
Following \cite{heger:04} we estimate the integrated deceleration
as
\beq
\Delta \Omega \approx 30.6\,\, \mu_{30}^{4/5}\, \Omega_3^{3/5}\,\,,
\eeq
where $\mu_{30} = B_{12} R_6^3$, $B_{12}$ is the surface field
in units of 10$^{12}$~G, and $R_6$ is the neutron star radius in
units of 10$^{6}$~cm. $\Omega_3$ is the neutron star angular velocity
in units of 10$^3$~rad~s$^{-1}$. For a 2~ms period neutron
star, hung-up with a 20~km radius and a conservative value of the
surface magnetic field strength of  
10$^{12}$~G, the integrated change in $\Omega$ is
$\sim$320~rad~s$^{-1}$, corresponding to a change in the period of
only 0.23~ms. With 15 times the magnetic field strength and all other
parameters kept fixed, the integrated change in $\Omega$ would
amount to $\sim$2800~rad~s$^{-1}$, leaving behind a 
neutron star with a period of $\sim$18~ms. 
Given these numbers, there is great potential for the propeller 
mechanism to lead to a significant, order-of-magnitude spindown 
of the neutron star. However, it is not clear whether the propeller 
mechanism works in the way suggested in the literature. Furthermore, 
there is general ignorance concerning the magnetic field strength 
and topology in neutron stars. Also, there may not be much, or any, 
fallback for the majority of supernovae.

\section{Summary and Discussion}
\label{section:sumdisc}

We have systematically  
investigated the mapping between presupernova 
(rotational) structure and the spin periods and rotational profiles 
of protoneutron stars.  For this, we have performed a large set of 
rotating supernova core-collapse simulations in spherical- and 
axi-symmetry, using Newtonian gravity. For spherically-symmetric runs, 
we have employed the Boltzmann radiation-hydrodynamics code SESAME
which has recently been upgraded to treat rotation in an approximate
way. In addition, viscous dissipation and angular momentum 
redistribution have 
been included in a subset of our simulations.
For models evolved in axisymmetry, we have made 
use of the multi-group flux-limited diffusion variant 
of the radiation-hydrodynamics code VULCAN/2D. 
For our simulations, we have taken 
11, 15, and 20 \mo\ (s11, s15, s20) presupernova
models from the stellar evolution study of \cite{ww:95} and put
them into rotation via a rotation law that assumes constant
angular velocity either on cylindrical or spherical shells. 
In addition, we have performed simulations with the progenitor models
E15A and E20A of \cite{heger:00} and m15b6 of \cite{heger:04}. 
These progenitor models include a one-dimensional 
prescription for rotational evolution. We have considered initial 
central iron core spin periods between 1.25~s and 157.1~s and  
only central periods that map to angular velocities below the Keplerian 
break-up limit at all radii in each model at all times during the numerical
evolution. The initial range of the rotation parameter $\beta$ is set by
a minimum of 7~$\times$~10$^{-5}$\% and a maximum of 0.88\%.

\begin{figure}
\begin{center}
\includegraphics[width=7.5cm]{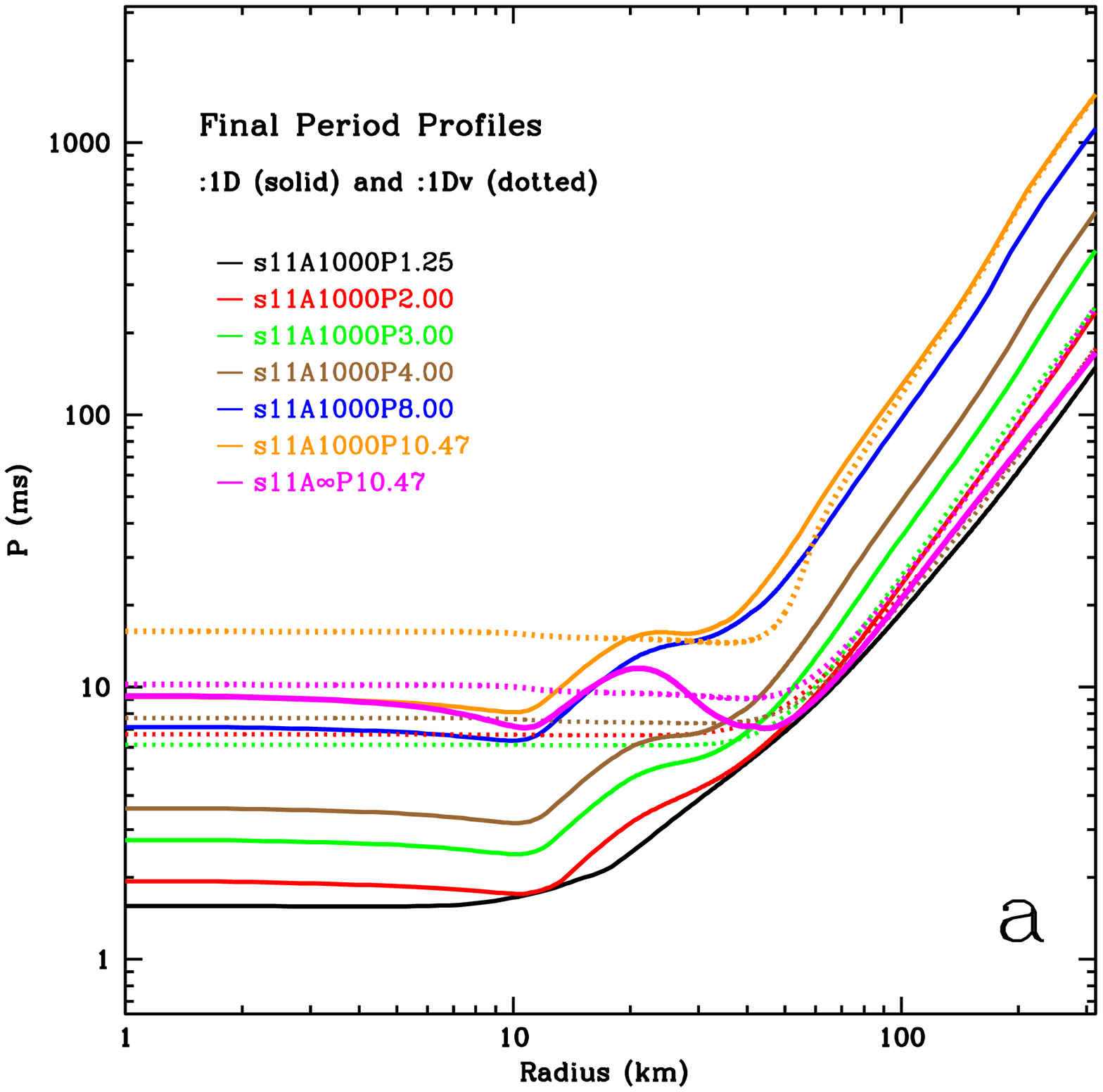} 
\includegraphics[width=7.5cm]{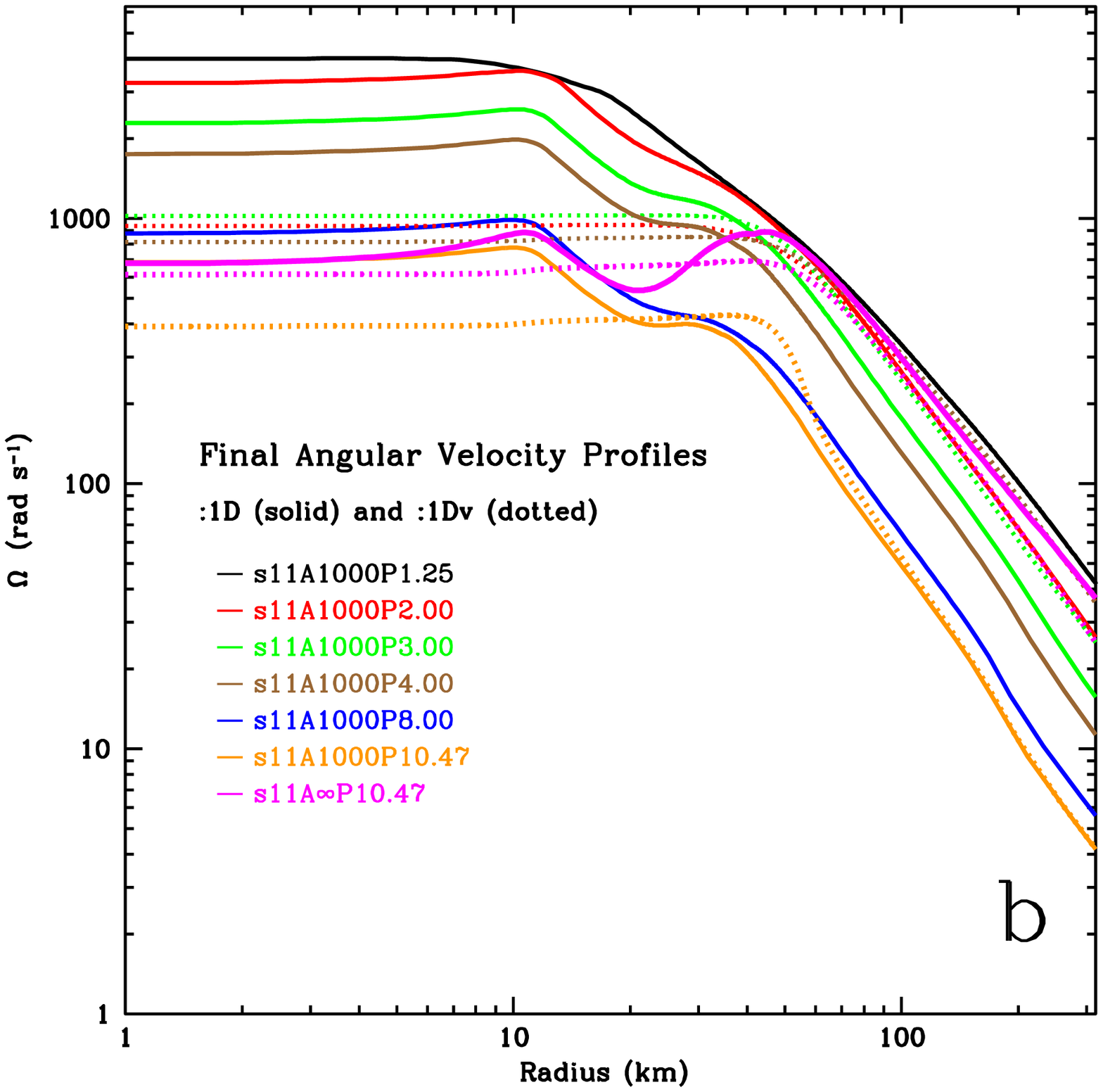} 
\caption{Profiles of the period (a) and the angular velocity (b)
at the end of each simulation for selected models from the 
s11 model series, evolved in spherical symmetry with SESAME. 
The models are initially forced into rotation
via the rotation law given in eq.~(\ref{eq:rotlaw}) with a choice
of $\Omega_0 = 2 \pi \mathrm{P_0}^{-1}$ and A as specified
in the model name. Model graphs plotted in solid lines correspond
to spherically symmetric evolution without the inclusion of 
viscous effects (:1D suffix
in Tables \ref{table:initialmodels} and \ref{table:results1D}). Model
graphs plotted in dotted lines correspond to spherically-symmetric 
evolution with viscous dissipation and angular momentum redistribution 
(:1Dv suffix in Tables \ref{table:initialmodels} 
and \ref{table:results1D}). For models
s11A1000P1.25 (black) and s11A1000P8.00 (blue), no :1Dv evolution 
is carried out. Note the approximate solid-body rotation of all :1D 
models out to $\sim$10 km.
Also note the factor of 3 to 4 larger region of solid-body rotation
realized by viscous dissipation in all :1Dv models. 
\label{fig:s11final}}
\end{center}
\end{figure}

\begin{figure}
\begin{center}
\includegraphics[width=7.5cm]{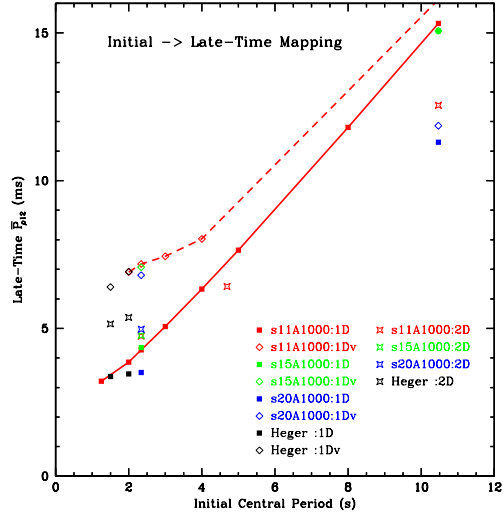} 
\caption{Mapping from the initial central iron core spin
period to the mean PNS period at 
$\sim$450~ms after bounce for the one-dimensional and $\sim$200~ms
after bounce for the two-dimensional models. We choose these times
for consistency among models from each model set. Most two-dimensional
models are evolved beyond $\sim$200~ms after bounce while most
one-dimensional models are evolved beyond $\sim$450~ms after bounce
 (see Tables~\ref{table:results1D}~and~\ref{table:results2D}).
On the abscissa, we plot the initial central
period in seconds and on the ordinate we plot the 
moment-of-inertia-weighted mean period of the PNS. 
Models evolved  with radiation-hydrodynamics in spherical 
symmetry are marked by solid boxes. Spherically-symmetric 
models that include viscosity are marked by open rhombi. Open
stars symbolize axisymmetric models. In addition, the model icons
are color-coded according to their progenitor type. s11 models are 
represented by red, s15 by green, and s20 by blue. Models from
the recent studies of \cite{heger:00} and \cite{heger:04} are marked
black. For the models considered in this study, rotational effects
are not dominant and the mapping between initial and final rotation
period is approximately linear.
\label{fig:initialfinalP}}
\end{center}
\end{figure}

In our one-dimensional simulations we find a roughly linear relationship
between initial central iron core spin period and the protoneutron
star spin. As an analysis tool we define a 
moment-of-inertia-weighted period (\pav) for all material above 
10$^{12}$~g~cm$^{-3}$. In Fig.~\ref{fig:initialfinalP}, we plot
\pav as a function of initial central spin period for a representative
subset of our models. Each one-dimensional model (:1D) is marked with
a solid box in the P$_0$ -- \pav plane. The boxes of our s11A1000:1D
model series are connected with solid lines to emphasize the
P$_0$ $\rightarrow$ \pav mapping. The graph is roughly fitted
by a linear function with slope of 1.3~ms~s$^{-1}$. Only in the
fastest models (initial periods shorter than $\sim$2.5~s) do centrifugal
effects lead to a noticable deviation from the displayed 
simple linear relationship. In considering the discrete 
profiles, we identify a region of near-solid body rotation inside
the nuclear-density region of the PNS out to
a radius of $\sim$10~km or $\sim$0.5-0.7~\mo\ in mass coordinate. 
This is a consequence of quasi-homologous
collapse. Outside this region PNSs are strongly 
differentially rotating. In a typical fast model, the angular 
velocity drops from $\sim$2500~rad~s$^{-1}$ at 20~km to a few hundreds
rad~s$^{-1}$ at 100~km. The energy stored in differential rotation 
can efficiently be tapped by viscous processes. We have considered the
MRI as one such process and have included its action in a subset
of our one-dimensional models (the ``:1Dv'' models; 
see \S\ref{section:1Dmethods}). As displayed 
in Fig.~\ref{fig:s11final},
viscosity leads to a spin down of the
central part of the PNS and enforces rigid
rotation out to 50~km and beyond. The viscous timescale
is proportional to $\Omega^{-1}$ and the 
viscous heating rate scales
with $\Omega^3$. Hence, the efficiency of viscosity in reducing
the rotational shear, in moving angular momentum out, and in 
spinning down the PNS is only great for rapidly rotating models.
In Fig.~\ref{fig:initialfinalP}, open rhombi symbolize the models
that include viscous angular momentum redistribution. Those rhombi
corresponding to models of the s11A1000:1D series are connected
with dashed lines. Comparing the two graphs in 
Fig.~\ref{fig:initialfinalP}, we note that the 
inclusion of viscosity leads to slower
\pav for all models. The relative spin down is greatest for models
below $\sim$3~ms period, but no model is slowed by more
than a factor of two in \pav. Models with long initial periods are slowed
by only a few to ten percent.

We have compared the one-dimensional models with our two-dimensional,
axisymmetric simulations that handle rotation consistently.
We find that the 1D approach correctly reproduces the overall 
phenomenology of the mapping between initial model and rotational
profile to postbounce PNS spin. However, we identify non-negligible
qualitative and quantitative differences when comparing individual
models evolved in 1D and 2D. In fast two-dimensional models convective motions
are confined to small cylindrical radii. Hence equatorial rotational
profiles of fast 2D models exhibit less substructure and are closer 
to their 1D counterparts than slower models in which convection 
is more vigorous out to larger radii.
For slow to moderate initial spin rate, the two-dimensional simulation 
results also yield a mapping of P$_0$ $\rightarrow$ \pav which is 
roughly linear for large P$_0$ and that deviates from linearity
at lower values of P$_0$; the correct treatment of centrifugal 
effects results in less compact and more oblate PNSs.

In comparing our 2D radiation-hydrodynamics, MGFLD models with purely 
hydrodynamic results, we find that the large-scale, coherent, and
repeated expansion-collapse-bounce cycles observed in 
purely hydrodynamic calculations are critically damped in
simulations that include neutrino radiative transfer and microphysics. 
This is an important finding because of its implication for the
gravitational wave signature of rotating stellar core collapse and 
bounce. Contrary to current belief (\citealt{zm:97,harry:02b,
kotake:03,ott:04}), the collapse and bounce
of even fast rotators (like E15A with P$_0$=1.5~s and 
an initial $\beta$ of 0.42\%) will not yield gravitational waves of 
large amplitude and energy emission. As with slowly rotating models,
the greatest part of the gravitational wave emission will come
from postbounce large-scale convective motions in the postshock
region and in the PNS (\citealt{ott:04}) and, possibly, 
from anisotropic neutrino emission (\citealt{bh:96,mueller:04}).

At the end of our two-dimensional simulations
(on average 240~ms after core bounce), we find mean PNS 
periods in the range from 4.7~ms to 140~ms. The final \pav depends
mainly on the inner precollapse iron core spin period. Differences
in progenitor mass, structure, and rotation law lead to only first-order
corrections in \pav. However, differences in progenitor mass and 
structure have a strong impact on other observables;
the more massive a progenitor, the greater are its 
final PNS mass, angular momentum, and rotation 
parameter $\beta$.
The maximum values of $\beta$ range between 5$\times$10$^{-2}$\% 
and 12.9\% in our two-dimensional simulations. At postbounce
times, $\beta$ increases roughly linearly as the PNS
simultaneously accretes more material and becomes more compact. 
During its cooling and deleptonization phase, a PNS
with a birth period below $\sim$5~ms will surpass the $\beta$-threshold 
values for classical secular and dynamical rotational instability 
($\sim$14\% and $\sim$27\%, respectively). It is, however, possible
that they experience a low-T/$|$W$|$ instability even earlier 
(\citealt{rotinst:05}). 

To compare with previous, though more approximate, 
studies, we have taken the PNS configurations of
our simulations at 200~ms after core bounce 
and extrapolated the spin periods and $\beta$s to final, 
deleptonized, and cold states. For the same (or very similar) 
initial conditions, our models  yield shorter neutron star periods 
than previous 
studies (\citealt{heger:00,fryerheger:00,
hegerwoosleyspruitlanger:03,heger:04}). Any model with initial iron
core period below $\sim$10~s leads to a neutron star with a 
submillisecond period, spinning with surface velocities close to 
break-up. 
Since young pulsars enter the $P$~--~$\dot{P}$ diagram at periods
of many tens to hundreds of milliseconds (\citealt{kaspihelfand:02}),
we have surveyed a number of possible spindown mechanisms that 
could transform a young millisecond-period neutron star into one  
with a period in agreement with observations.
These include viscous angular momentum
redistribution by the MRI or another viscous process, rotational
hydrodynamic instabilities, r-modes, neutrino angular momentum
transport, rotation-powered explosions, early magnetic-dipole
(pulsar) emission, neutrino-driven magnetic winds, and the
propeller mechanism. We find that none
of the processes considered is robust enough to spin down a millisecond
PNS to a tens-of-milliseconds pulsar. The most promising
is the propeller mechanism that removes angular momentum from the 
neutron star by expelling fall-back material at the Alfv\'en 
radius, while not requiring magnetic field strengths in excess of 
several~10$^{12}$~G to a few 10$^{13}$~G. 
However, there remain many open questions concerning 
the physical details involved and whether fallback 
is even generic has not been resolved.

Hence, we conclude that, if a pulsar were born rapidly rotating,
nonaxisymmetric rotational hydrodynamic instabilities
and other proprosed mechanisms of spindown may not be  
adequate to cause the requisite spindown 
from rapid to observed rates.  In addition, it is clear that
spindown mechanisms that employ magnetic fields would
also transfer a rapidly rotating PNS's kinetic energy to
the envelope debris, in contrast with the observational estimates of
generic supernova and supernova remnant energies (\S\ref{magneto}).
Therefore, on the basis of our theoretical calculations of the
mapping between initial and final spins, likely spindown 
mechanisms, and the observational constraints from pulsars
and supernova energetics, we conclude that the progenitor
spin rates of the cores of the massive stars that give birth
to observed neutron stars and pulsars are low, with
corresponding periods on average greater than $\sim$50 
seconds.  The associated average neutron star would thereby 
be born with a period no less than $\sim$10~ms and with $\beta$~\sless~2\%. 
However, the final neutron star spin is 
a function of the progenitor mass. Some very massive
progenitors might lead to very fast and massive neutron stars that
might evolve to collapsars and rotation-powered GRBs. 

\begin{figure}
\begin{center}
\includegraphics[width=7.5cm]{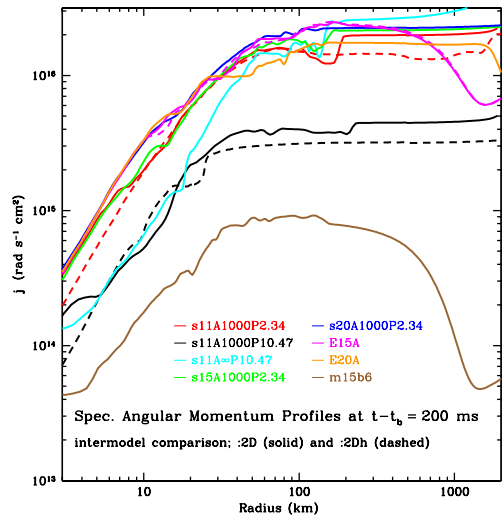} 
\caption{Equatorial specific angular momentum ($j$) profiles at
200~ms after core bounce plotted for a representative 
set of models evolved in axisymmetry with VULCAN/2D. Drawn in
solid lines are graphs of models evolved with MGFLD and dashed lines
represent models that are evolved purely hydrodynamically and
undergo prompt explosions. At 200~ms after bounce, the PNS 
is already relatively compact and most of the mass on the grid is 
located interior to $\sim$50~km (Fig.~\ref{fig:intermodel_2D_mass}).
Models that make use of the E15A, E20A (\citealt{heger:00}), and 
m15b6 (\citealt{heger:04}) presupernova models show drops in
their $j$ profiles at large radii which correspond to the discontinuities
in their precollapse rotational profiles (Fig.~\ref{fig:initialrot}).
\label{fig:intermodel_2D_j}}
\end{center}
\end{figure}

Figure~\ref{fig:intermodel_2D_j} shows equatorial specific angular
momentum profiles at 200~ms after core bounce for a representative 
subset of our models evolved in axisymmetry with VULCAN/2D. Note that
models of similar initial central iron core P$_0$ yield very similar
specific angular velocity profiles, despite differences in progenitor
structure. Model s11A$\infty$P10.47:2D, which is initially set up in 
solid-body rotation, yields outside some tens of kilometers 
specific angular momenta as high as found in differentially 
rotating models with up to 5~times shorter initial central periods. 
Given the estimates of \cite{hegerwoosley:03}, the specific angular 
momenta of our models with initial central periods below $\sim$10~s 
are sufficient to  rotationally support an accretion disk around a 
Kerr black hole.

Finally, we point out that even though this study has 
advanced our knowledge concerning the birth spin periods and rotational
profiles of neutron stars, and their dependence on initial
iron core spin and structure, many important questions remain
unanswered. We have followed the PNSs
for only a few hundred milliseconds after formation. We have not
tracked the supernova explosion through the stellar envelope
and subsequent fall-back accretion.
We have not included general relativistic effects, nor 
magneto-hydrodynamics. Fully consistent, three-dimensional, 
radiation-magneto-hydrodynamics simulations in general relativity 
will be required to reveal in detail the evolution from 
a very young PNS to the final rotating pulsar 
(or black hole).

\acknowledgments

We acknowledge discussions with and help from Jeremiah Murphy, 
Itamar Lichtenstadt, Casey Meakin, and Ed Seidel. 
Importantly, we acknowledge support for this work
from the Scientific Discovery through Advanced Computing 
(SciDAC) program of the DOE, grant number DE-FC02-01ER41184,
and from the NSF, grant number AST-0504947. 
E. L. acknowledges support from the Israel Science
Foundation under grant 805/04. R.W. thanks 
the Institute of Astronomy of the ETH Z\"urich for providing
part-time office space. T.A.T. acknowledges support by NASA 
through Hubble Fellowship grant \#HST-HF-01157.01-A, awarded by 
the Space Telescope Science Institute, which is operated by the 
Association of Universities for Research in Astronomy, Inc., for 
NASA, under contract NAS 5-26555. Finally, we thank Jeff Fookson 
and Neal Lauver of the Steward Computer Support Group for their 
invaluable help with the local Beowulf clusters and acknowledge 
the use of the NERSC/LBNL/seaborg and  ORNL/CCS/RAM machines. 
Movies and additional still frames associated 
with this work can be obtained upon request from the authors.


\clearpage

\begin{deluxetable}{ccccllccr}

\tabletypesize{\scriptsize}
\tablewidth{17cm}
\tablecaption{Initial Models\label{table:initialmodels}}
\tablehead{
\colhead{Progenitor}&
\colhead{ZAMS Mass}&
\colhead{R$_{Fe}$}&
\colhead{M$_{Fe}$}&
\colhead{Model Name}&
\colhead{Evolution}&
\colhead{A}&
\colhead{$\Omega_0$}&
\colhead{$P_0$}\\
\colhead{}&
\colhead{(\mo)}&
\colhead{(10$^8$ cm)}&
\colhead{(\mo)}&
\colhead{}&
\colhead{Type}&
\colhead{(km)}&
\colhead{(rad s$^{-1}$)}&
\colhead{(s)}
}
\startdata
s11&11&1.35&1.32&&&&\\
&&&&s11A1000P1.25:1D\tablenotemark{a}&SESAME&1000&5.03&1.25\\
&&&&s11A1000P2.00:1D\tablenotemark{a}&SESAME&1000&3.14&2.00\\
&&&&s11A1000P2.34:1D&SESAME&1000&2.68&2.34\\
&&&&s11A1000P3.00:1D\tablenotemark{a}&SESAME&1000&2.09&3.00\\
&&&&s11A1000P4.00:1D\tablenotemark{a}&SESAME&1000&1.57&4.00\\
&&&&s11A1000P5.00:1D\tablenotemark{a}&SESAME&1000&1.26&5.00\\
&&&&s11A1000P8.00:1D\tablenotemark{a}&SESAME&1000&0.79&8.00\\
&&&&s11A1000P10.47:1D&SESAME&1000&0.60&10.47\\

&&&&s11A1000P2.00:1Dv\tablenotemark{a}&SESAME V&1000&3.14&2.00\\
&&&&s11A1000P2.34:1Dv&SESAME V&1000&2.68&2.34\\
&&&&s11A1000P3.00:1Dv\tablenotemark{a}&SESAME V&1000&2.09&3.00\\
&&&&s11A1000P4.00:1Dv\tablenotemark{a}&SESAME V&1000&1.57&4.00\\
&&&&s11A1000P10.47:1Dv&SESAME V&1000&0.60&10.47\\
&&&&s11A$\infty$P10.47:1D&SESAME&$\infty$&0.60&10.47\\
&&&&s11A$\infty$P10.47:1Dv&SESAME V&$\infty$&0.60&10.47\\
&&&&s11A1000P2.34:2D\tablenotemark{b}&VULCAN&1000&2.68&2.34\\
&&&&s11A1000P2.34:2Dh&VULCAN H&1000&2.68&2.34\\
&&&&s11A1000P4.69:2D\tablenotemark{b}&VULCAN&1000&1.34&4.69\\
&&&&s11A1000P10.47:2D\tablenotemark{b}&VULCAN&1000&0.60&10.47\\
&&&&s11A1000P10.47:2Dh&VULCAN H&1000&0.60&10.47\\
&&&&s11A1000P41.89:2D\tablenotemark{b}&VULCAN&1000&0.15&41.89\\
&&&&s11A1000P157.10:2D\tablenotemark{b}&VULCAN&1000&0.04&157.10\\
&&&&s11A$\infty$P10.47:2D&VULCAN&$\infty$&0.60&10.47\\

s15&15&1.16&1.28&&&&\\
&&&&s15A1000P2.34:1D&SESAME&1000&2.68&2.34\\
&&&&s15A1000P10.47:1D&SESAME&1000&0.60&10.47\\
&&&&s15A1000P2.34:1Dv&SESAME V&1000&2.68&2.34\\
&&&&s15A1000P10.47:1Dv&SESAME V&1000&0.60&10.47\\
&&&&s15A1000P2.34:2D&VULCAN&1000&2.68&2.34\\
s20&20&2.21&1.92&&&&\\
&&&&s20A1000P2.34:1D&SESAME&1000&2.68&2.34\\
&&&&s20A1000P10.47:1D&SESAME&1000&0.60&10.47\\
&&&&s20A1000P2.34:1Dv&SESAME V&1000&2.68&2.34\\
&&&&s20A1000P10.47:1Dv&SESAME V&1000&0.60&10.47\\
&&&&s20A1000P2.34:2D&VULCAN&1000&2.68&2.34\\
E15A&15&2.22&1.62&&&&\\
&&&&E15A:1D\tablenotemark{a}&SESAME&-&4.18&1.50\\
&&&&E15A:1Dv\tablenotemark{a}&SESAME V&-&4.18&1.50\\
&&&&E15A:2D&VULCAN&-&4.18&1.50\\
&&&&E15A:2Ds&VULCAN S&-&4.18&1.50\\
E20A&20&2.78&1.84&&&&\\
&&&&E20A:1D&SESAME&-&3.13&2.00\\
&&&&E20A:1Dv&SESAME V&-&3.13&2.00\\
&&&&E20A:2D&VULCAN&-&3.13&2.00\\
m15b6&15&1.95&1.47&&&&\\
&&&&m15b6:1D&SESAME&-&0.198&31.73\\
&&&&m15b6:1Dv&SESAME V&-&0.198&31.73\\
&&&&m15b6:2D&VULCAN&-&0.198&31.73\\
\enddata
\tablecomments{
\ List of models included in this study. The supernova 
progenitor star models s11, s15 and s20 are taken from \cite{ww:95}; 
E15A and E20A are 
from \cite{heger:00} and m15b6 is from \cite{heger:04}. Models 
E15A, E20A, and m15b6 include rotation using a one-dimensional
prescription for angular momentum evolution. Centrifugal force 
terms were included 
up to the end of core carbon burning. s11, s15, and s20 are non-rotating
and are forced to rotate by the rotation law given in 
eq.~(\ref{eq:rotlaw}). $R_{Fe}$ is the radius of 
the iron core (determined by the discontinuity in $Y_e$ at the 
outer edge of the iron core). $M_{Fe}$ is the mass of the iron core.  
Evolution type denotes the way a model is evolved. SESAME stands for 
spherically symmetric evolution with the SESAME code, SESAME V indicates 
that dissipative  angular momentum redistribution is switched on, 
VULCAN means two-dimensional simulation with MGFLD, VULCAN S 
stands for two-dimensional simulation with MGFLD, but \it shellular \rm
rotation (\S\ref{section:2Dmethods}), and VULCAN H means
purely hydrodynamic two-dimensional simulation without neutrinos. 
A is the parameter 
governing the degree of differential rotation  
(eq.~\ref{eq:rotlaw}). $\Omega_0$ and P$_0$ are the initial central 
angular velocity and spin period, respectively. 
All progenitor models already have an initial infall 
velocity profile when they are mapped onto our computational
grid. Nothing artificial is done to initiate collapse.
}
\tablenotetext{a}{Previously published/discussed in \cite{thompvisc:05}}
\tablenotetext{b}{Previously published/discussed in \cite{walder:05}}
\end{deluxetable}
\clearpage

\begin{deluxetable}{lrrrrrcl}
\tablewidth{17.5cm}
\tabletypesize{\scriptsize}
\tablecaption{Summary of Results: One-Dimensional Models
\label{table:results1D}}
\tablehead{
\colhead{Model Name}&
\colhead{t$_f$ - t$_b$}&
\colhead{P$_{0}$}&
\colhead{$\bar{\mathrm{P}}_{\rho 12,f}$}&
\colhead{$M_{\rho 12,f}$}&
\colhead{$J_{\rho 12,f}$}&
\colhead{$\beta_{i}$}&
\colhead{$\beta_{f}$}\\
\colhead{}&
\colhead{(ms)}&
\colhead{(s)}&
\colhead{(ms)}&
\colhead{(\mo)}&
\colhead{(10$^{49}$ erg s)}&
\colhead{(\%)}&
\colhead{(\%)}
}
\startdata
s11A1000P1.25:1D   
&587&1.25&3.11&1.07&1.172&8.80$\cdot$10$^{-1}$&18.44\\

s11A1000P2.00:1D   
&629&2.00&3.50&1.28&1.127&3.44$\cdot$10$^{-1}$&10.45\\

s11A1000P2.34:1D   
&630&2.34&3.85&1.32&1.091&2.51$\cdot$10$^{-1}$
&\phantom{0}7.96\\

s11A1000P3.00:1D   
&588&3.00&4.72&1.35&0.901&1.53$\cdot$10$^{-1}$
&\phantom{0}5.10\\

s11A1000P4.00:1D   
&541&4.00&5.94&1.37&0.706&8.60$\cdot$10$^{-2}$
&\phantom{0}2.98\\

s11A1000P5.00:1D   
&608&5.00&6.94&1.39&0.586&5.50$\cdot$10$^{-2}$
&\phantom{0}1.97\\

s11A1000P8.00:1D
&532&8.00&11.16&1.39&0.366&2.15$\cdot$10$^{-2}$
&\phantom{0}0.759\\

s11A1000P10.47:1D  
&549&10.47&14.28&1.39&0.283&1.25$\cdot$10$^{-2}$
&\phantom{0}0.448\\

\hline

s11A1000P2.00:1Dv  
&500&2.00&6.72&1.13&0.392&3.44$\cdot$10$^{-1}$
&\phantom{0}2.25\\

s11A1000P2.34:1Dv
&437&2.34&7.18&1.16&0.428&2.51$\cdot$10$^{-1}$
&\phantom{0}2.42\\

s11A1000P3.00:1Dv  
&1037&3.00&6.13&1.29&0.500&1.53$\cdot$10$^{-1}$
&\phantom{0}1.97\\

s11A1000P4.00:1Dv  
&573&4.00&7.42&1.33&0.524&8.60$\cdot$10$^{-2}$
&\phantom{0}1.99\\

s11A1000P10.47:1Dv  
&549&10.47&14.93&1.39&0.271&1.25$\cdot$10$^{-2}$
&\phantom{0}0.416\\

\hline
\hline

s11A$\infty$P10.47:1D
&571&10.47&9.18&1.34&0.428&5.51$\cdot$10$^{-2}$
&\phantom{0}1.59\\

s11A$\infty$P10.47:1Dv
&570&10.47&9.41&1.34&0.415&5.51$\cdot$10$^{-2}$
&\phantom{0}1.17\\

\hline
\hline

s15A1000P2.34:1D
&612&2.34&3.95&1.39&1.197&2.45$\cdot$10$^{-1}$
&\phantom{0}7.92          \\

s15A1000P10.47:1D
&562&10.47&13.70&1.47&0.316&1.22$\cdot$10$^{-2}$
&\phantom{0}0.468       \\

s15A1000P2.34:1Dv
&612&2.34&6.76&1.24&0.501&2.45$\cdot$10$^{-1}$
&\phantom{0}2.51          \\

s15A1000P10.47:1Dv
&563&10.47&14.13&1.46&0.307&1.22$\cdot$10$^{-2}$
&\phantom{0}0.447      \\

\hline
\hline

s20A1000P2.34:1D
&462&2.34&3.48&1.84&2.820&3.53$\cdot$10$^{-1}$
&15.26\\

s20A1000P10.47:1D
&349&10.47&11.30&1.85&0.645&1.76$\cdot$10$^{-2}$
&\phantom{0}0.875        \\

s20A1000P2.34:1Dv
&365&2.34&6.80&1.40&0.749&3.53$\cdot$10$^{-1}$
&\phantom{0}4.32          \\

s20A1000P10.47:1Dv
&347&10.47&11.86&1.85&0.620&1.76$\cdot$10$^{-2}$
&\phantom{0}0.740       \\
\hline

E15A:1D                
&780&1.50&2.91&1.53&2.064&4.95$\cdot$10$^{-1}$
&17.45                \\

E15A:1Dv               
&639&1.50&5.76&1.25&0.513&4.95$\cdot$10$^{-1}$
&\phantom{0}2.50      \\

\hline
\hline

E20A:1D                
&461&2.00&3.43&1.84&2.213&2.85$\cdot$10$^{-1}$&12.52\\

E20A:1Dv               
&361&2.00&6.92&1.46&0.805&2.85$\cdot$10$^{-1}$
&\phantom{0}3.95      \\

\hline
\hline

m15b6:1D
&589&31.73&57.98&1.50&0.077&7.77$\cdot$10$^{-4}$
&\phantom{0}0.0280   \\

m15b6:1Dv              
&589&31.73&58.21&1.50&0.076&7.77$\cdot$10$^{-4}$
&\phantom{0}0.0234   \\
\enddata               

\tablecomments{Numerical results for the supernova simulations performed 
with SESAME in spherical symmetry. Model initial conditions are 
detailed in Table \ref{table:initialmodels}. Here, t$_f$ - t$_b$ 
denotes the time elapsed after core bounce to when we stop the 
calculation. P$_{0}$ is the initial central rotation period 
and $\bar{\mathrm{P}}_{\rho\mathrm{12},f}$ is the mean 
PNS rotation period (for material with 
$\rho~\ge$~10$^{12}$~g~cm$^{-3}$ (defined by 
eq.~\ref{eq:pavrho}) at the end of the simulation.
$M_{\rho\mathrm{12},f}$ is the integrated mass of all material above the
same density threshold at the end of the numerical calculation 
and $J_{\rho\mathrm{12},f}$ is the angular momentum associated with $M_{\rho\mathrm{12},f}$. 
$\beta_i$ and $\beta_f$ represent the 
values of T/$|$W$|$ at the initial stage and at the end of the 
simulation, respectively.
}
\end{deluxetable}

\begin{deluxetable}{lrrrrrcl}
\tablewidth{17.5cm}
\tabletypesize{\scriptsize}
\tablecaption{Summary of Results: Two-Dimensional Models
\label{table:results2D}}
\tablehead{
\colhead{Model Name}&
\colhead{t$_f$ - t$_b$}&
\colhead{P$_{0}$}&
\colhead{$\bar{\mathrm{P}}_{\rho 12,f}$}&
\colhead{$M_{\rho 12,f}$}&
\colhead{$J_{\rho 12,f}$}&
\colhead{$\beta_{i}$}&
\colhead{$\beta_{f}$}\\
\colhead{}&
\colhead{(ms)}&
\colhead{(s)}&
\colhead{(ms)}&
\colhead{(\mo)}&
\colhead{(10$^{49}$ erg s)}&
\colhead{(\%)}&
\colhead{(\%)}
}
\startdata
s11A1000P2.34:2D
&197
&2.34&4.74&1.17&1.522&3.14$\cdot$10$^{-1}$
&\phantom{0}6.72\\

s11A1000P2.34:2Dh
&150&2.34&4.86&1.16&1.436&3.14$\cdot$10$^{-1}$
&\phantom{0}6.12\\

s11A1000P4.69:2D&
262&4.69&5.63&1.28&0.998&8.35$\cdot$10$^{-2}$
&\phantom{0}2.64\\

s11A1000P10.47:2D
&293&10.47&10.05&1.33&0.489&1.53$\cdot$10$^{-2}$
&\phantom{0}0.694       \\

s11A1000P10.47:2Dh
&294&10.47&10.44&1.13&0.301&1.53$\cdot$10$^{-2}$
&\phantom{0}0.498        \\

s11A1000P41.89:2D
&319&41.89&37.79&1.33&0.123&1.04$\cdot$10$^{-3}$
&\phantom{0}0.0536      \\

s11A1000P157.10:2D
&318&157.10&139.63&1.33&0.033&7.00$\cdot$10$^{-5}$
&\phantom{0}0.00451   \\

\hline
\hline

s11A$\infty$P10.47:2D
&192&10.47&9.50&1.26&0.613&5.35$\cdot$10$^{-2}$
&\phantom{0}1.26      \\

\hline
\hline

s15A1000P2.34:2D
&215&2.34&4.69&1.18&1.563&3.07$\cdot$10$^{-1}$
&\phantom{0}7.00          \\

\hline
\hline

s20A1000P2.34:2D   
&238&2.34&4.98&1.42&3.351&5.08$\cdot$10$^{-1}$
&12.67                    \\

\hline
\hline

E15A:2D                
&217&1.50&5.12&1.28&2.801&6.45$\cdot$10$^{-1}$
&12.85\\

E15A:2Ds
&243&1.50&4.73&1.37&2.738&4.64$\cdot$10$^{-1}$
&10.75\\

\hline
\hline

E20A:2D                
&211&2.00&5.40&1.49&3.185&4.17$\cdot$10$^{-1}$
&10.76                \\

\hline
\hline

m15b6:2D               
&195&31.73&50.25&1.39&0.129&1.03$\cdot$10$^{-3}$
&\phantom{0}0.0533   \\
\enddata               

\tablecomments{
Same as Table~\ref{table:results1D}, but listing results of
our simulations performed with
VULCAN/2D in axisymmetry.}
\end{deluxetable}

\begin{deluxetable}{lrrrrrr}
\tabletypesize{\scriptsize}
\tablecaption{Final Neutron Star Spin Estimates
\label{table:simpleP}}
\tablehead{
\colhead{Model Name}&
\colhead{P$_{0}$}&
\colhead{$M_{\rho 12,200}$}&
\colhead{$J_{\rho 12,200}$}&
\colhead{P$_{NS}$}&
\colhead{$\beta_{NS}$}&
\colhead{$\Omega_{NS} / \Omega_{K}$}\\
\colhead{}&
\colhead{(s)}&
\colhead{(\mo)}&
\colhead{(10$^{49}$ erg s)}&
\colhead{(ms)}&
\colhead{(\%)}&
\colhead{}
}
\startdata
s11A1000P2.34:2D
&2.34
&1.17
&1.522
&0.48
&50.65
&1.37\\

s11A1000P4.69:2D
&4.69
&1.25
&0.992
&0.85
&15.16
&0.75\\

s11A1000P10.47:2D
&10.47
&1.28
&0.443
&1.82
&3.25
&0.35\\

s11A1000P41.89:2D
&41.89
&1.29
&0.113
&7.19
&0.21
&0.09\\

s11A1000P157.10:2D
&157.10
&1.29
&0.030
&27.10
&0.02
&0.02\\

s11A$\infty$P10.47:2D
&10.47
&1.26
&0.613
&1.30
&6.54
&0.49\\

s15A1000P2.34:2D
&2.34
&1.18
&1.550
&0.48
&51.25
&1.38\\

s20A1000P2.34:2D   
&2.34
&1.27
&3.098
&0.28
&129.01
&2.20\\

E15A:2D                
&1.50
&1.27
&2.724
&0.29
&126.05
&2.17\\

E15A:2Ds
&1.50
&1.33
&2.553
&0.33
&96.01
&1.89\\

E20A:2D                
&2.00
&1.47
&3.104
&0.30
&104.13
&1.98\\

m15b6:2D               
&31.73
&1.39
&0.129
&6.79
&0.21
&0.09\\
\hline
\hline
s11A1000P2.34:1D
&2.34
&1.15
&0.762
&0.95
&13.39
&0.70\\

s11A1000P2.34:1Dv
&2.34
&1.06
&0.378
&1.77
&4.23
&0.39\\

s11A1000P10.47:1D
&10.47
&1.26
&0.211
&3.76
&0.77
&0.17\\

s11A1000P10.47:1Dv
&10.47
&1.25
&0.200
&3.94
&0.71
&0.16\\

s11A$\infty$P10.47:1D
&10.47
&1.23
&0.295
&2.63
&1.63
&0.25\\

s11A$\infty$P10.47:1Dv
&10.47
&1.22
&0.275
&2.80
&1.45
&0.23\\

s15A1000P2.34:1D
&2.34
&1.19
&0.790
&0.95
&12.96
&0.69\\

s15A1000P2.34:1Dv
&2.34
&1.09
&0.416
&1.65
&4.71
&0.42\\

s20A1000P2.34:1D
&2.34
&1.41
&1.622
&0.55
&32.35
&1.10\\

s20A1000P2.34:1Dv
&2.34
&1.15
&0.527
&1.38
&6.41
&0.49\\

E15A:1D
&1.5
&1.21
&1.271
&0.60
&31.86
&1.09\\

E15A:1Dv
&1.5
&1.01
&0.378
&1.68
&4.91
&0.42\\

E20A:1D
&2.00
&1.49
&1.465
&0.64
&22.24
&0.92\\

E20A:1Dv
&2.00
&1.19
&0.560
&1.34
&6.51
&0.49\\

m15b6:1D
&31.37
&1.33
&0.059
&14.21
&0.05
&0.04\\

m15b6:1Dv
&31.37
&1.33
&0.058
&14.45
&0.05
&0.04\\

\enddata               

\tablecomments{Overview table listing the key quantities
for the cold neutron star spin estimates of \S\ref{section:simpleP}.
P$_{0}$ is the initial central 
rotation period, $M_{\rho\mathrm{12},200}$ is the mass of all material 
with density $\rho \ge$~10$^{12}$~g~cm$^{-3}$ at 200~ms after bounce, 
and $J_{\rho\mathrm{12},200}$ is the
total angular momentum contained in that material. $P_{NS}$ is the 
final neutron star spin estimate according to eq. (\ref{eq:simpleP}). 
$\beta_{NS}$ is an estimate of the neutron star rotation parameter and
$\Omega_{NS}/\Omega_{K}$ is the ratio of the neutron star's
angular velocity to the Keplerian break-up velocity defined by 
$\sqrt{GM/r^3}$. Neutron stars with 
$\Omega_{NS}/\Omega_{K}$~\sgreat~1.0 or $\beta$~\sgreat~50\% 
cannot exist.}
\end{deluxetable}

\clearpage

\clearpage

\clearpage

\end{document}